\documentclass[pre,amsmath,amssymb,showpacs,preprint]{revtex4}

\usepackage{graphicx}
\usepackage{bm}
\def\be{\begin{equation}}
\def\ee{\end{equation}}
\def\A{\tilde A}
\def\ps{\tilde\psi}
\def\L{\tilde L}
\def\AL{\tilde\alpha}
\def\R{\tilde R}
\def\t{\tilde\tau}
\def\ep{\epsilon}
\def\sal{\sigma_{\rm AL}}
\def\nab{\bm{\nabla}}
\newcommand{\REAL}{{\sf R\hspace*{-1.02ex}\vspace*{-0.1ex}%
  \rule{0.14ex}{1.4ex}\hspace*{1.02ex}\vspace*{0.1ex}}}
\def\xiq{\tilde\xi^2}

\begin{document}
\title{Ginzburg--Landau equations with consistent Langevin terms for nonuniform wires}
\author{Jorge Berger}
\affiliation{Physics Department, Ort Braude College, P. O. Box 78,
21982 Karmiel, Israel 
}
\email{jorge.berger@braude.ac.il}
\begin{abstract}
Many analyses based on the time-dependent Ginzburg--Landau model are not consistent with statistical mechanics, because thermal fluctuations are not taken correctly into account. We use the fluctuation-dissipation theorem in order to establish the appropriate size of the Langevin terms, and thus ensure the required consistency. Fluctuations of the electromagnetic potential are essential, even when we evaluate quantities that do not depend directly on it. Our method can be cast in gauge-invariant form.
We perform numerous tests, and all the results are in agreement with statistical mechanics. We apply our method to evaluate paraconductivity of a superconducting wire. The Aslamazov--Larkin result is recovered as a limiting situation. Our method is numerically stable and the nonlinear term is easily included. We attempt a comparison between our numerical results and the available experimental data. Within an appropriate range of currents, phase slips occur, but we found no evidence for thermally activated phase slips. We studied the behavior of a moderate constriction. A constriction pins and enhances the occurrence of phase slips.
\end{abstract}
\maketitle
\section{INTRODUCTION}
Almost a century ago Langevin \cite{Lang} proposed a ``complementary" stochastic force in order to enable the description of an individual Brownian particle in terms of Newton's second law. This approach gave origin to the field of stochastic differential equations.

Among the many widely used stochastic differential equations, we will focus on the time-dependent Ginzburg--Landau equations (TDGL), which are used to describe a superconductor out of equilibrium \cite{Sch1}. The TDGL equations may be written in the form
\be
\gamma\hbar{\partial_t \psi }=-[\alpha+\beta|\psi |^2+|\alpha|\xi^2(i\nab -2\pi{\bf A}/\Phi_0)^2]\psi \;, 
\label{psi} 
\ee
\be
\sigma {\partial_t {\bf A}}=(2\hbar c e/m){\rm Re} [ {\psi^*}
( i\nab -2\pi{\bf A}/\Phi_0) \psi ] -(c^2/4\pi)\nab \times \nab \times {\bf A}\;,  
\label{ohm}
\ee
where $\psi $ is the order parameter, ${\bf A}$ the electromagnetic vector potential, $t$ the time, $\xi$ the coherence length, $\Phi_0=hc/2e$ the quantum of magnetic flux, $\alpha $, $\beta $ and $\gamma $ are material parameters, $\sigma$ is the normal electrical conductivity, $c$ is the speed of light, $e$ the absolute value of the electron charge, $m$ the mass of a Cooper pair and the asterisk denotes complex conjugation. $\beta $ and $\gamma $ are positive, whereas $\alpha $ is positive above the critical temperature and negative below it. The product $|\alpha|\xi^2=\hbar ^2/2m$ does not depend on the temperature. In these equations the gauge choice eliminates the scalar potential.

In order to cope with fluctuation problems, Schmid added a Langevin term to these equations \cite{Sch2}. Precisely, he added to the right-hand side of Eq.~(\ref{psi}) a random function of position and time, $f({\bf r},t)$, such that
\be
\langle f^*({\bf r},t) f({\bf r'},t')\rangle =2\Omega \hbar\gamma k_B T \delta ({\bf r}-{\bf r'})\delta (t-t') \;,
\label{deltas}
\ee
where $\langle\cdots\rangle$ denotes ensemble average, $\Omega $ is the volume of the sample, $k_B$ is Boltzmann's constant and $T$ is the temperature. Since then the use of a Langevin term has become almost a standard practice in the numeric solution of the TDGL equations; some miscellaneous examples are \cite{Kato,KatoC,Bolech,Boris,Tarlie,charge,HP}. Generalizations of Eq.~(\ref{deltas}) to various classes of models were reviewed in \cite{HH}.

We shall be particularly interested in wires with nonuniform cross section. This interest is motivated by previous works \cite{berger,nikulov,giles}, which found that nonuniform superconducting wires may behave qualitatively differently than a uniform wire, especially near the transition temperature. We are therefore interested in a numeric scheme that can enable us to investigate nonuniform wires in the region in which thermal fluctuations are important.  
For this purpose we would like to integrate the TDGL equations by means of Euler-Maruyama iterations. As we will discuss after Eq.~(\ref{algo}), the existing literature does not provide an explicit procedure for evaluating the size of the fluctuations of the order parameter in each computational site; in particular, it is not mentioned that this size depends on the volumes of the computational cells.

Moreover, addition of a Langevin term to Eq.~(\ref{ohm}) \cite{HP,Dorsey,Machida}, which can take into account fluctuations of the electromagnetic field, is the exception rather than the rule. We shall see that this omission is not self-consistent, even if only the values of the order parameter are of our concern. We are unaware of the initial motivation for including fluctuations in the order parameter only; a belated argument could be the result \cite{FsqH}
that charges of an isotropic superfluid are irrelevant for $T\xi\alt 1\,$K$\,$cm, but this result does not refer to the 1D situation that we consider.  

An early review on fluctuations near superconducting phase transitions was written by Skocpol and Tinkham \cite{ST}. The voltage-current characteristic in quasi-one-dimensional superconductors was extensively studied \cite{LAMH,oldexp,IofE,skocpol,Likharev,Barat,Maslova,Rangel} and reviewed \cite{Ivlev,Tidecks} several decades ago. The most salient feature is the presence of voltage steps, which are atributed to ``phase slips" in which the order parameter vanishes at a point and releases a winding from its phase. In recent years there has been a renewed interest in this topic \cite{Pb,ac,ph-slip,S-shape,MichotteB69,koby}, fuelled by the availability of long nanoscopic superconducting wires. Among other features, it has been possible to investigate the existence of quantum phase slips and the behavior under fixed voltage.

In this article we build the equations for the evolution of a nonuniform superconducting wire, as appropriate for a finite difference numeric treatment. For this purpose we first discretize the TDGL thermodynamic potential and then invoke the fluctuation-dissipation theorem \cite{FDT} to evaluate the self-consistent Langevin terms.

Our method is developed in Sec.~\ref{METHOD}. Expressions for the noise term at each computational element are obtained and we find a convenient way to determine how the Johnson noise affects the order parameter. Section \ref{TOY} compares the results of our method with those of statistical mechanics for simple situations in which analytic results can be obtained. In Sec.~\ref{APPL} our method is used for situations in which analytic results are not available. Appendix \ref{AA} provides numerical details that contributed to the success of our algorithm.

\section{OUR METHOD\label{METHOD}}
\subsection{The fluctuation-dissipation theorem}
For our purposes, the theorem may be stated as follows. We consider a system with energy $G$ that depends on several variables, one of which is $x$. Let the system be in thermal contact with a heat bath at temperature $T$ and let $x$ be a classical variable, i.e., a real function of time. Let $x$ obey on the average the macroscopic equation
\be
dx/dt=-\Gamma \partial G/\partial x \;.
\label{mac}
\ee
Let $\tau$ be a macroscopically short period of time.
Then the microscopic change of $x$, $x(t+\tau)-x(t)$, will be $-\Gamma\tau \partial G/\partial x$ on the average and its
variance will be
\be
\langle \eta ^2\rangle =2\Gamma k_BT\tau \;,
\label{FDvar}
\ee
where $\eta =x(t+\tau)-x(t)+\Gamma\tau \partial G/\partial x$.

The physical idea behind this theorem is the same as in the original Langevin article, i.e., the same interactions of the variable $x$ with the heat bath, that give origin to its microscopic fluctuations, are those that lead to its macroscopic evolution in Eq.~(\ref{mac}).

Several assumptions are necessary for the validity of this theorem. One of them is that the distribution of the coordinates of the heat bath that interact with $x$ remain described by the canonical ensemble. There are also conditions on the size of $\tau$; it has to be macroscopically small, quantitatively expressed by $\tau \ll k_BT/\Gamma (\partial G/\partial x)^2$ and $\tau \ll 1/\Gamma \partial^2 G/\partial x^2$. On the other hand, $\tau $ has to be microscopically large, explicitly, much larger than the autocorrelation time of the Langevin term and also large enough to allow for a classical treatment of $x$. From the numerical point of view, taking a value of $\tau$ that is too small may be a waste of resources but, in principle, should not affect the results. Indeed, if we take a time step $\tau /K$ rather than $\tau $, the average and the variance of the change in $x$ will decrease by the factor $K$, so that, if $\eta $ has Gaussian distribution, after $K$ steps the ensemble of results will be the same as if the time step were $\tau$.

\subsection{Ginzburg-Landau energy}
In this model the energy of a superconducting wire can be written as
\be
G=\int [\alpha|\psi |^2+\beta|\psi |^4/2+|\alpha|\xi^2|(i\nab -2\pi{\bf A}/\Phi_0)\psi|^2+IA/cw]dV \;,
\label{GGL}
\ee
where the integral is over the volume of the wire. Here $I$ is the current that flows along the wire, $w$ is its cross section, and the last term in the integrand takes into account the energy provided by the power source. We consider a wire with a cross section that is much smaller than the square of the magnetic penetration depth, so that the induced magnetic field can be neglected. In this situation we are also allowed to take the electromagnetic potential ${\bf A}$ in the direction of the wire. Likewise, we will assume that the width of the wire is small compared to the coherence length and varies slowly with the arc length; for the superconducting-insulator boundary condition these assumptions imply that $\nab $ may be replaced with the derivative with respect to the arc length.

Let us now discretize this energy. We denote the length of the wire by $L$ and divide it into $N$ segments of length $L/N$. Each segment has to be sufficiently short, so that the order parameter, the elecromagnetic potential and the cross section may be considered to remain uniform within it, and we denote their values by $\psi_k$, $A_k$ and $w_k$ for the $k^{\rm th}$ segment. Defining the grid-dependent dimensionless variables $\tilde\xi=N\xi/L$ and  $\A =2\pi LA/N\Phi_0$, we may write
\begin{eqnarray}
G&=&\frac{L}{N}\sum w_k\left\{\alpha|\psi_k |^2+\beta|\psi_k|^4/2+ \right. \nonumber\\
&&\left. 
\frac{|\alpha|}{2}\xiq \left[ |(1+i\A_k)\psi_k-\psi_{k-1}|^2+|(-1+i\A_k)\psi_k+\psi_{k+1}|^2\right]\right\}+
\frac{I\Phi_0}{2\pi c}\sum\A_k \;.
\label{discrete}
\end{eqnarray}
The intuitive meaning of $\tilde\xi$ is ``the number of consecutive segments over which the order parameter remains approximately uniform." In numeric calculations $\tilde\xi$ has to be of the order of 1; if $\tilde\xi$ is too small the discretized model does not represent the physical sample and, if it is too big, the calculation becomes inefficient. 

Discretization of the local terms in Eq.~(\ref{GGL}) is quite natural, but discretization of the term with the gradient is somewhat arbitrary. 
Since the term $(i\nab -2\pi{\bf A}/\Phi_0)\psi$ requires the values of the vector potential and of the derivative of $\psi$ at the same site, it is usually convenient to use a staggered approach, in which ${\bf A}$ and $\psi$ are defined at alternating overlapping sites. However, since in Eq.~(\ref{GGL}) $|(i\nab -2\pi{\bf A}/\Phi_0)\psi|^2$ is multiplied by the cross section of the wire, which may not be uniform, the use of overlapping sites is problematic.
It is tempting to replace in Eq.~(\ref{discrete}) the term in square brackets with $|\psi_{k+1}-\psi_{k-1}+2i\A_k \psi_k|^2/2$. Unfortunately, even in the limit $N\gg 1$, this form would lead to two separate sublattices ($k$ odd and $k$ even) such that there is no penalty for changing their relative phases; this feature does not represent the original continuous model.

We decompose $\psi_k$ into real variables $\psi_k=u_k+iv_k$.
The derivative of the discretized energy with respect to $u_k$ may be written as
\begin{eqnarray}
\frac{\partial G}{\partial u_k}&=&\frac{Lw_k}{N}{\rm Re}\left\{2\alpha \psi_k+2\beta|\psi_k|^2\psi_k+
|\alpha| \xiq\left[2(1+\A_k^2)\psi_k-(1+i\A_k)\psi_{k+1}-(1-i\A_k)\psi_{k-1}+ \right. \right. \nonumber\\
&&\left. \left. w_k^{-1}[w_{k-1}(\psi_k-(1-i\A_{k-1})\psi_{k-1})+w_{k+1}(\psi_k-(1+i\A_{k+1})\psi_{k+1})]
\right]\right\} \;.
\label{derivu}
\end{eqnarray}
The derivative with respect to $v_k$ involves the imaginary part of the same expression, which is actually $2\partial G/\partial\psi_k^*$.

The derivative of the energy with respect to $\A_k$ is
\be
\frac{\partial G}{\partial \A_k}=\frac{I\Phi_0}{2\pi c}+\frac{N|\alpha|\xi^2 w_k}{L}
[(v_{k+1}-v_{k-1}+2\A_k u_k)u_k-(u_{k+1}-u_{k-1}-2\A_k v_k)v_k] \;.
\label{dera}
\ee
Noting that the term in square brackets is proportional to a discretized version of the superconducting current density and substituting $|\alpha|\xi^2=\hbar ^2/2m$, Eq.~(\ref{dera}) reduces to
\be
\frac{\partial G}{\partial \A_k}=\frac{I_N\Phi_0}{2\pi c} \;,
\label{IN} 
\ee
where $I_N$ is the normal current along the wire.

\subsection{Evolution equations}
Comparison of Ohm's law with Eq.~(\ref{IN}) gives the macroscopic equation
\be
d\A_k/dt=-\Gamma_{A,k} \partial G/\partial \A_k \;,
\label{macA}
\ee
with
\be
\Gamma_{A,k}=\frac{4e^2L}{N\hbar^2\sigma w_k} \;.
\label{GamA}
\ee

Equation (\ref{derivu}) has to be compared with the evolution of the order parameter in a wire \cite{giles}. Using superconductor-insulator boundary conditions and our notation, this evolution is given by
\be
\gamma\hbar{\partial_t \psi }=-\{\alpha+\beta|\psi |^2+|\alpha|\xi^2[(i\partial_s -2\pi A/\Phi_0)^2+i(w'(s)/w(s))(i\partial_s -2\pi A/\Phi_0)]\}\psi \;,
\label{gil}
\ee
where $s$ denotes the arc length. Assuming that $w$, $\psi$ and $A$ can be differentiated twice, Eq.~(\ref{gil}) is obtained as the limit of 
\be
du_k/dt=-\Gamma_{\psi,k} \partial G/\partial u_k \;, 
\label{macu}
\ee
and an analogous equation for $v_k$, with
\be
\Gamma_{\psi,k}=\frac{N}{2\gamma \hbar Lw_k} \;.
\label{Gamu}
\ee
The macroscopic evolution of $u_k$ and $v_k$ can be described by the complex equation $d\psi_k/dt=-2\Gamma_{\psi,k} \partial G/\partial\psi_k^*$.

We close this section with a summary of our basic numerical algorithm. Let $\psi_k$ and $\A_k$ be known at a certain instant. Then, after a time step $\tau$ their new values will be
\begin{eqnarray}
\psi_k&\leftarrow &\psi_k-2\Gamma_{\psi,k}\tau\frac{\partial G}{\partial \psi_k^*}+\eta_u+i\eta_v \nonumber \\
\A_k&\leftarrow &\A_k-\Gamma_{A,k}\tau\frac{\partial G}{\partial \A_k}+\eta_A \;,
\label{algo}
\end{eqnarray}
where $\eta_u$, $\eta_v$ and $\eta_A$ are random numbers with gaussian distribution, zero average and variances 
$\langle \eta_u^2\rangle =\langle \eta_v^2\rangle =2\Gamma_{\psi,k} k_BT\tau$, $\langle \eta_A^2\rangle =2\Gamma_{A,k} k_BT\tau$, with the derivatives of $G$ given by Eqs. (\ref{derivu})--(\ref{dera}) and the $\Gamma$'s given by Eqs. (\ref{GamA}) and (\ref{Gamu}). It should be emphasized that $\Gamma_{A,k}$ and $\Gamma_{\psi,k}$ depend on the length and the cross section of the $k^{\rm th}$ segment.

In order to use algorithm (\ref{algo}), $\langle \eta_{u,v,A}^2\rangle$ must be known. Nevertheless, practically all the literature limits itself to mentioning the continuum-style relation (\ref{deltas}). Neither the expressions nor the procedure for evaluating $\langle \eta_{u,v,A}^2\rangle$ are provided. To our knowledge, the only exceptions are \cite{Kato} and \cite{KatoC}. The expressions provided in these references neither coincide with those we have obtained nor among themselves. According to us, the expression in Eq.~(16) of \cite{Kato} should vanish for a trully infinitely long sample; expression (18) in \cite{KatoC} seems to depend on the volume of the entire sample rather than the volume of the computational cell. Reference \cite{Bolech} just quotes \cite{Kato} and \cite{KatoC} (although the size of the noise is essentially regarded as a free computational parameter), and \cite{charge} relies on \cite{Bolech}. Many studies (e.g. \cite{Tarlie}) do not require the variance of $\eta_{u,v,A}$, because algorithm (\ref{algo}) is actually not used. A possibility for avoiding use of (\ref{algo}) is working in Fourier space, but this approach is not useful for a nonuniform wire.

\subsection{A gauge-invariant method\label{GI}}
For a one dimensional wire the electromagnetic potential can be gauged out of the evolution equation by means of an appropriate transformation. Defining a gauge function 
\be
{\cal U}(s)=\exp\left(\frac{2\pi i}{\Phi_0}\int_0^s A(s')ds'\right)
\label{U}
\ee
and the gauge-invariant order parameter $\ps (s)={\cal U}(s)\psi (s)$, the term $|(i\nab -2\pi{\bf A}/\Phi_0)\psi|$ in Eq.~(\ref{GGL}) reduces to $|d\ps /ds|$. We discretize ${\cal U}$ as
\be
{\cal U}_k=\exp\left(i\sum_{j=1}^{k-1/2}\A_j\right) \;,
\label{discreteU}
\ee
where the sum up to a half-integer number is understood to contain the term $\A_k$, but multiplied by $\frac{1}{2}$. Once ${\cal U}_k$ is defined, we can write $\ps_k={\cal U}_k\psi_k$. A discretized version of the energy can then be written as
\begin{eqnarray}
G&=&\frac{L}{N}\sum \left\{w_k(\alpha|\ps_k |^2+\beta|\ps_k|^4/2)+ \right. \nonumber\\
&&\left. 
\frac{|\alpha|}{2}\xiq (w_k+w_{k+1})|\ps_{k+1}-\ps_k|^2\right\}+
\frac{I\Phi_0}{2\pi c}\sum\A_k \;,
\label{GGLtil}
\end{eqnarray}
where $w_k$ is understood as 0 for $k=0$ and $k=N+1$.

Although $\ps_k$ is physically more meaningful than $\psi_k$ ($|d\ps /ds|$ is proportional to the velocity), it is not a ``canonical" variable, i.e., $d\ps_k/dt\neq -2\Gamma\partial G/\partial \ps_k^*$. Instead, we have
\be
\frac{d\ps_k}{dt}={\cal U}_k\frac{d\psi_k}{dt}+i\ps_k \sum_{j=1}^{k-1/2}\frac{d\A_j}{dt} \;.
\label{29}
\ee
Noting that 
\be
\frac{d\psi_k}{dt}=-2\Gamma_{\psi,k}\frac{\partial G}{\partial \ps_k^*}\frac{\partial \ps_k^*}{\partial \psi_k^*} \;,
\label{30}
\ee
the first term in Eq.~(\ref{29}) takes the value $-2\Gamma_{\psi,k}\partial G/\partial \ps_k^*$, i.e., its contribution is what we would obtain if $\ps_k$ were a canonical variable. In order to obtain the evolution of $\A_j$, we express the energy in Eq.~(\ref{GGLtil}) in terms of $\psi$ rather than $\ps$ and then use Eq.~(\ref{macA}). This gives
\be
\frac{d\A_j}{dt}=-\Gamma_{A,j}\left(\frac{I\Phi_0}{2\pi c}+\frac{|\alpha|L\xiq}{2N}{\rm Im}\left[(w_{j-1}+w_{j})\ps_{j-1}^*\ps_j+(w_{j}+w_{j+1})\ps_{j}^*\ps_{j+1}\right]\right) \;,
\label{32}
\ee
which is the discrete version of Ohm's law required by Eq.~(\ref{GGLtil}). Substituting these results into Eq.~(\ref{29}) we obtain the macroscopic evolution of $\ps$. This evolution can be followed without any knowledge about $\A$. 

We investigate now the influence of fluctuations. $\ps_k$ is the same as $\psi_k$, rotated in the complex plane; therefore, its fluctuations arise from those of $\psi_k$ and from those in the angle of rotation. Since the size of the fluctuations of the real and the imaginary part of $\psi_k$ are the same and since they are not correlated, their contribution to the fluctuations of $\ps_k$ has the same distribution as $\eta_u+i\eta_v$. The influence of the fluctuations in the angle of rotation can be taken into account as follows. When integrating the evolution of $\ps_k$ in Eq.~(\ref{29}) we require the macroscopic change of $\A_j$, $\tau d\A_j/dt$. To this macroscopic change, we have to add the fluctuation of $\A_j$ during the period $\tau $, $\eta_A$, which is still described by $\langle \eta_A^2\rangle =2\Gamma_{A,j} k_BT\tau$.

Our algorithm for the evolution of $\ps_k$ consisted of compound steps. The first stage of a step, that corresponds to the first term in Eq.~(\ref{29}), looks the same as the evolution of $\psi_k$ in Eq.~(\ref{algo}), with $A_k$ set to zero. The second stage takes care of the last term in Eq.~(\ref{29}); instead of a standard Euler step we multiply by a phase, i.e.,
\be
\ps_k\leftarrow\ps_k\prod_{j=1}^{k-1/2}\exp\left(i\frac{\A_j}{dt}\tau+i\eta_{A,j}\right) \;
\label{algotil}
\ee
where the meaning of the upper limit is that the argument of the exponential for $j=k$ is divided by 2. To first order in $\tau$ and $\eta_{A,j}$ this method is equivalent to an Euler step, but, since it keeps $|\ps_k|$ unchanged, is numerically more stable. 

\subsection{Dimensionless parameters}
Besides $\tilde\xi$ and $\A$, that were defined before Eq.~(\ref{discrete}), it will be convenient to define the following dimensionless quantities. $\AL=\alpha/k_BT$ provides a conveniently normalized value of the condensation energy. $\L=L(2mk_BT)^{1/2}/\hbar$ is the length of the wire divided by the ``thermal wavelength," except for a factor $\sqrt{\pi}$. We also define $\R=4e^2\gamma L/\hbar\sigma  w_0$, where $w_0$ is some typical cross section of the wire. For a wire of uniform cross section, $\R$ is the normal resistance of the wire multiplied by $2\pi\gamma$ and divided by the quantum of resistance, $h/(2e)^2$. $\R$ is proportional to the ratio between the relaxation times of the order parameter and the electromagnetic potential. $\t=\alpha\tau/\gamma \hbar$ is the effective size of the iteration steps; it is not dictated by the physics of the problem and we expect that, provided that $\t$ is significantly smaller than 1 and significantly larger than the inverse of the number of iterations, the results should not depend on it. Some additional quantities, which are useful when $\beta \neq 0$, will be defined in Sec.~\ref{Num}.

\section{TOY MODELS\label{TOY}}
In order to test our method, we would like to compare its results against quantities that can be evaluated exactly. For this purpose, the following sections consider simplified models.
\subsection{Uniform $\A$}
We drop the last term in Eq.~(\ref{discrete}), that causes the energy to be unbounded from below, and also set $\beta =0$, so that the problem becomes linear in the order parameter. In order to keep $G$ bounded from below, we must set $\alpha >0$. As additional simplifications we will consider a uniform wire $w_k=w_0$ for $k=1,\dots ,N$, and a uniform value $\A_k=\A$ for all the segments. With these simplifications the energy becomes
\be
G=\frac{\alpha Lw_0}{N}\sum \left\{|\psi_k|^2+ \frac{\xiq}{2} \left[ |(1+i\A)\psi_k-\psi_{k-1}|^2+|(-1+i\A)\psi_k+\psi_{k+1}|^2\right] \right\} \;.
\label{simple}
\ee
We also have to specify boundary conditions at the extremes of the wire. Let us consider periodic boundary conditions, $\psi_0=\psi_N$, $\psi_{N+1}=\psi_1$. 

For a fixed value of $\A$, $G$ can be diagonalized by a Fourier transformation. We define $\varphi_n=N^{-1}\sum\psi_k e^{-2\pi i kn/N}$ and, for simplicity of notation, we will take even values of $N$. The inverse transformation can be written as
\be
\psi_k=\sum_{n=-N/2+1}^{N/2}\varphi_n e^{2\pi i kn/N} \;.
\label{inverse}
\ee
Substituting Eq.~(\ref{inverse}) into Eq.~(\ref{simple}) we obtain
\be
G=\alpha Lw_0\sum_{n=-N/2+1}^{N/2}|\varphi_n|^2\left(1+\xiq [(\A+\sin 2\pi n/N)^2+4\sin ^4\pi n/N]\right) \;.
\label{diagonal}
\ee

The ensemble average of any quantity $Q$ is given by
\be
\langle Q\rangle =\int Q(\{\varphi_n\};\A)e^{-G/k_BT}d^2\varphi_{-N/2+1}\cdots d^2\varphi_{N/2}d\A/\int e^{-G/k_BT} d^2\varphi_{-N/2+1}\cdots d^2\varphi_{N/2}d\A \;,
\label{ensemble}
\ee
where $\A$ varies over $\REAL$ and $\varphi_n$ over $\REAL^2$. If $Q$ is a polynomial, the integrals over $\varphi_n$ are easily evaluated. For example, the denominator in Eq.~(\ref{ensemble}) (the partition function) is
\be
Z=\int_{-\infty}^\infty \frac{(\pi k_BT/\alpha Lw_0)^N d\A}{(1+\xiq (\A^2+4))\prod_{n=1-N/2}^{N/2-1}(1+\xiq [(\A+\sin 2\pi n/N)^2+4\sin ^4\pi n/N])} \;.
\label{Z}
\ee
This last integral may be either evaluated numerically or collecting the residues at $\A=\sin 2\pi n/N+i(\tilde\xi^{-2}+4\sin^4\pi n/N)^{1/2}$.

Note that, since the dependence on $\A$ factors out in the integrand of Eq.~(\ref{Z}), it contributes just a multiplicative constant to the partition function. Therefore, $\A$ is a degree of freedom that makes no contribution to the heat capacity. For the same reason, the expectation value for any function of $\A$ is independent of the temperature. 

In our calculations using algorithm (\ref{algo}) we started from $u_k=v_k=\A=0$ and let these variables build up from fluctuations. We performed $N_{\rm relax}$ iterations to achieve a ``typical" situation and then averaged the quantities of interest during $N_{\rm av}$ additional iterations. 

\begin{figure}
\scalebox{0.85}{\includegraphics{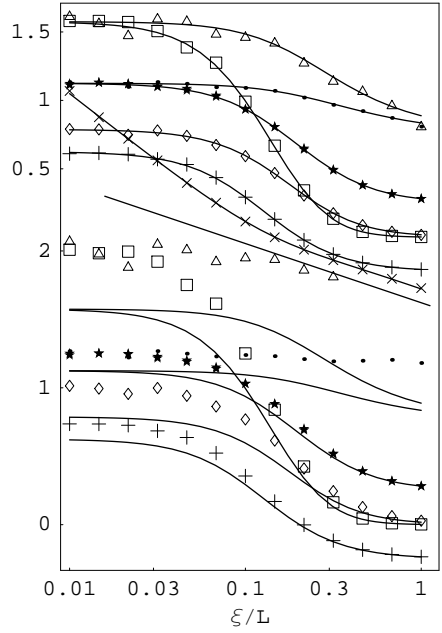}}%
\caption{\label{psin4}Averages of several quantities, as functions of the coherence length. The symbols correspond to values calculated using our method (upper part) and the lines are ensemble averages. The parameters used in the calculation were $N=4$, $\t=3\times 10^{-5}$, $\R=16000$, $N_{\rm av}=2\times 10^8$, $N_{\rm relax}=6\times 10^7$. The parameter $\AL$ factors out. $\bullet$ $\AL Lw_0\langle|\varphi_0|^2\rangle+0.25$ (the constant 0.25 was added for visibility); $\star$ $\AL Lw_0\langle|\varphi_{-1}|^2\rangle+0.25$ and $\AL Lw_0\langle|\varphi_1|^2\rangle+0.25$; $+$ $\AL Lw_0\langle|\varphi_2|^2\rangle-0.25$; $\triangle$ $(\AL Lw_0)^2\langle|\varphi_0|^4\rangle$; $\Box$ $(\AL Lw_0)^2\langle|\varphi_1|^4\rangle$; $\Diamond$ $(\AL Lw_0)^2\langle|\varphi_0|^2|\varphi_1|^2\rangle$; $\times$ $0.5\log_{10}\langle\A^2\rangle$. Upper part: fluctuations of $\A$ were included; lower part: fluctuations of $\A$ were ignored. We have drawn an oblique straight line that separates the ``upper" and ``lower" parts.}
\end{figure}

Figure~\ref{psin4} compares the averages of several quantities, as obtained with our method for $N=4$, with those expected from Eq.~(\ref{ensemble}), for a wide range of values of $\xi/L$. If the fluctuations of $\A$ are ignored, we obtain the results in the lower part. As expected, ignoring the fluctuations of $\A$ yields a variance of $\A$ (not shown in the graph) that is much lower than its ensemble value, but we see in the lower part that also the averages of quantities that do not directly involve $\A$ turn out to be incorrect. 

\begin{figure}
\scalebox{0.85}{\includegraphics{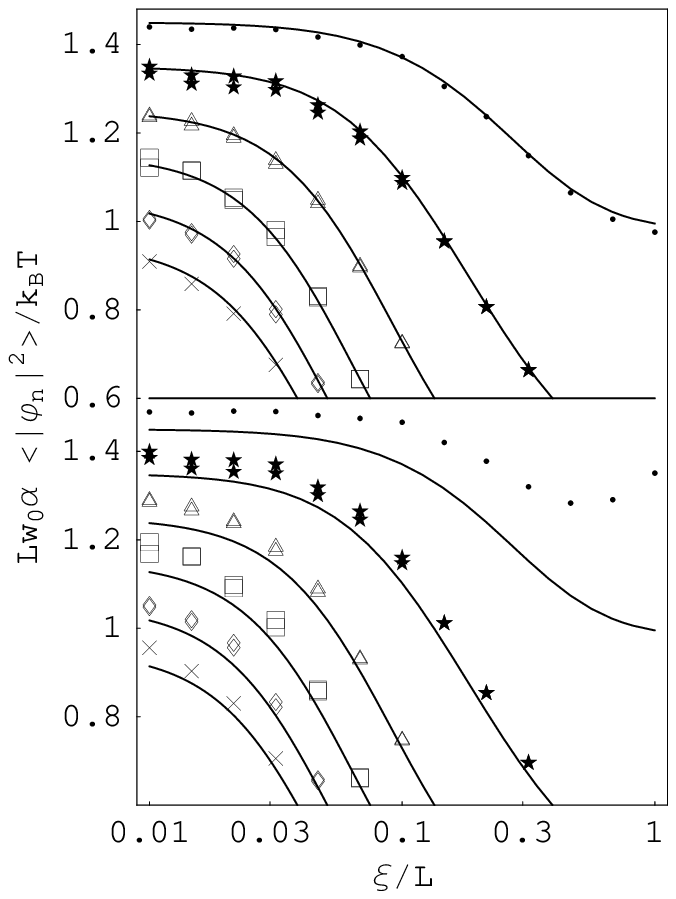}}%
\caption{\label{psin16}Variances of the Fourier components of the order parameter, as functions of the coherence length, for the case of 10 segments. The curve for $n=5$ sits at its true position; for visibility, the other curves have been raised by $(5-|n|)/10$. Besides $N$, the parameters 
are the same as in Fig.~\ref{psin4}. $\bullet$ $n=0$; $\star$ $n=\pm 1$; $\triangle$ $n=\pm 2$; $\Box$ $n=\pm 3$; $\Diamond$ $n=\pm 4$; $\times$ $n=5$. Upper panel: fluctuations of $\A$ included; lower panel: fluctuations of $\A$ ignored. For visibility, only the region above 0.6 is shown.}
\end{figure}

The upper panel in Fig.~\ref{psin16} compares the averages $\langle|\varphi_n|^2\rangle$ obtained with our method with those expected from Eq.~(\ref{ensemble}) for $N=10$,  and the lower panel repeats the comparison while ignoring fluctuations of $\A$. We see again that the symbols in the lower panel deviate from Eq.~(\ref{ensemble}), but this time the disagreement is smaller than in the case $N=4$. The reason is that the variable $\A$ is driven by the other variables $u_i$ and $v_i$; when $\xi/L$ is sufficiently large, the variance of $\A$ is practically independent of whether or not its fluctuations are considered. This is a feature of our toy model, in which there are $2N$ degrees of freedom for the order parameter and only one for $\A$; in the original model with $N$ degrees of freedom for $\A_i$ the influence of the fluctuations of the electromagnetic field should not decrease with $N$.

\subsection{\label{gautoy}Gauge-invariant formalism}
We still take $I=\beta =0$ and uniform cross section $w_k=w_0$, but use of the formalism developed in Sec.~\ref{GI} enables us to release the requirement of uniform $\A$. We consider two boundary conditions. One of them is $\ps(0)=\ps(L)=0$, which corresponds to contacts with a conductor that strongly suppresses superconductivity. In a discrete model, these conditions are approximated by $\ps_0=-\ps_1$ and $\ps_{N+1}=-\ps_N$. In this case we write
\be
\ps_k=\sum_{n=1}^N\varphi_n^s\sin\frac{n(k-1/2)\pi}{N}
\label{phis}
\ee
with
\be
\varphi_n^s=\frac{2-\delta_{nN}}{N}\sum_{k=1}^N\ps_k \sin\frac{n(k-1/2)\pi}{N} \;,
\label{phisdef}
\ee
where $\delta $ is Kroneker's symbol. Expression~(\ref{GGLtil}) then assumes the diagonal form
\be
G=\alpha Lw_0\sum_{n=1}^N(1+\delta_{nN})|\varphi_n^s|^2\left(\frac{1}{2}+2\xiq\sin^2\frac{n\pi}{2N}\right) \;.
\label{GGLs}
\ee

The other boundary condition we consider is that in which $d\ps /ds$ vanishes, which corresponds to contacts with an insulator. In a discrete model, we require $\ps_0=\ps_1$ and $\ps_{N+1}=\ps_N$. In this case we write
\be
\ps_k=\sum_{n=0}^{N-1}\varphi_n^c\cos\frac{n(k-1/2)\pi}{N}
\label{phic}
\ee
with
\be
\varphi_n^c=\frac{2-\delta_{n0}}{N}\sum_{k=1}^N\ps_k \cos\frac{n(k-1/2)\pi}{N} 
\label{phicdef}
\ee
and Eq.~(\ref{GGLtil}) reduces to
\be
G=\alpha Lw_0\left\{|\varphi_0^c|^2+\sum_{n=1}^{N-1}|\varphi_n^c|^2\left(\frac{1}{2}+2\xiq\sin^2\frac{n\pi}{2N}\right)\right\} \;.
\label{GGLc}
\ee

\begin{figure}
\scalebox{0.85}{\includegraphics{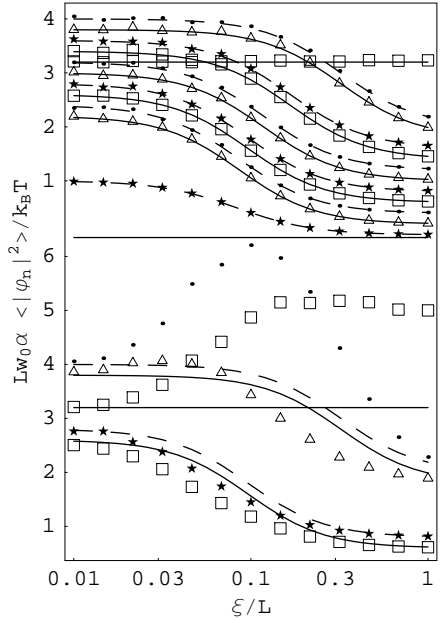}}%
\caption{\label{gaug}Variances of the Fourier components $\varphi_n^s$ and $\varphi_n^c$ of the gauge-invariant order parameter $\ps$, for $N=6$. 
The dashing lines describe $\langle |\varphi_n^s|^2 \rangle$ according to Eq.~(\ref{GGLs}) ($\ps =0$ at the boundaries), and have been raised by $0.4(6-n)$. The solid lines describe $\langle |\varphi_n^c|^2 \rangle$ according to Eq.~(\ref{GGLc}) ($d\ps/ds =0$ at the boundaries), and have been raised by $0.4(5.5-n)$. Note that $\langle |\varphi_0^c|^2 \rangle$ does not depend on $\xi$. The symbols describe the values obtained using the method developed in Sec. \ref{GI}. $\bullet$ $\langle |\varphi_n^s|^2 \rangle$, $n$ odd; $\star$ $\langle |\varphi_n^s|^2 \rangle$, $n$ even; $\triangle$ $\langle |\varphi_n^c|^2 \rangle$, $n$ odd; $\Box$ $\langle |\varphi_n^c|^2 \rangle$, $n$ even. Besides $N=6$ and $\t=10^{-5}$, the parameters are the same as in Fig.~\ref{psin4}. Upper panel: fluctuations of $\A$ included; lower panel: fluctuations of $\A$ ignored. In order to avoid clutter, only the cases $n=0$, $n=1$ and $n=4$ are shown in the lower panel.}
\end{figure}


Figure \ref{gaug} shows values of $\langle |\varphi_n^s|^2 \rangle$ and $\langle |\varphi_n^c|^2 \rangle$ obtained from Eqs. (\ref{phisdef}) and (\ref{phicdef}) after following the evolution of $\ps$ according to Sec. \ref{GI}. We see that, in spite of the fact that the vector potential $\A_k$ has been gauged out and the macroscopic evolution of $\ps$ can be described in closed form by means of Eqs. (\ref{29}) and (\ref{32}), the fluctuations of $\A_k$ still remain essential.

Inspection of the lower panel in Fig.~\ref{gaug} (including 7 additional curves that are not shown) suggests the following trends: (i) omission of the fluctuations of $\A$ yields values of $\langle |\varphi_n|^2 \rangle$ that are larger than the true values for small $n$ and smaller than the true values for large $n$, i.e., the order parameter appears to be flatter than it should; (ii) the largest deviations of $\langle |\varphi_n|^2 \rangle$ from its true values occur for coherence lengths of the order of the length of the wire; for higher harmonics, the largest deviation tends to occur at smaller coherence length.

Trend (i) might be understood if we consider the mechanism by which fluctuations of $\A_k$ influence the order parameter. For a bulk superconductor, the currents generated by these fluctuations induce magnetic fields that are felt by the order parameter; this is the mechanism considered, e.g., in \cite{golden}. For a very thin wire, the induced magnetic field is negligible. However, since the total current has to be uniform along the wire, the Johnson current forces a superconducting counter-current, which in turn forces a phase gradient. Therefore, fluctuations of $\A_k$ force the order parameter to be rougher than it would be in their absence. Note that at a very small scale, which is beyond the scope of our model, electroneutrality is not a realistic assumption. Trend (ii) may be qualitatively understood, since higher harmonics probe shorter lengths.

\subsection{\label{continuous}The continuous limit}
We consider now the limit $N\rightarrow\infty$. In this limit $\Gamma_A\rightarrow 0$, and we might expect that the fluctuations of $\A_k$ become unimportant. As in the previous section, we drop the nonlinear term, take a uniform cross section, and use the gauge-invariant order parameter $\ps$. This time we will consider the periodic boundary condition $\ps_{N+1}=\ps_1$, which physically corresponds to a ring that encloses a magnetic flux that equals an integer number of quanta. The case of a ring that encloses a non-integer flux will be considered elsewhere. For this boundary condition we cannot take $I=0$; instead, $I$ becomes a Lagrange multiplier. Numerically, the effect of this current amounts to the substraction of a uniform term from every $\A_k$, so that $\sum \A_k(t)$ remains constant. Since $\tilde\xi\propto N$, $d\ps/dt$ diverges in the limit $N\rightarrow\infty$; in order to overcome this difficulty we used the method described in Appendix \ref{external}.

Defining $\tilde\varphi_n=N^{-1}\sum\ps_k e^{-2\pi i kn/N}$, the statistical average of $\langle|\tilde\varphi_n|^2\rangle$ can be obtained by setting $\tilde A\rightarrow 0$ in Eq.~(\ref{diagonal}). From here we obtain $\langle|\tilde\varphi_n|^2\rangle=k_BT/\alpha Lw_0(1+4\xiq\sin^2(n\pi/N))$. Table~\ref{bigN} compares this statistical average with the values obtained with and without fluctuations of $\A_k$ for several values of $N$ and $n$. For the parameters considered, we see that the larger the number of elements into which we divide the sample, the more severe the error of the results obtained when the fluctuations of $\A_k$ are ignored. 

\begin{table*}
\caption{\label{bigN}$10^4\AL Lw_0\langle|\tilde\varphi_n|^2\rangle$ for several values of the number of elements $N$ and the wave number $n$. 
We have taken a coherence length equal to the length of the sample, and the other parameters are as in Fig.~\ref{psin4}.  }
\begin{ruledtabular}
\begin{tabular}{l|ccc|ccc|ccc|ccc}
 &\multicolumn{3}{c|}{$N=8$}&\multicolumn{3}{c|}{$N=16$}&\multicolumn{3}{c|}{$N=32$}&\multicolumn{3}{c}{$N=64$}\\ \hline
 $n$&\footnotemark[1]&\footnotemark[2]&\footnotemark[3]&\footnotemark[1]&\footnotemark[2]&\footnotemark[3]&\footnotemark[1]&\footnotemark[2]&\footnotemark[3]&\footnotemark[1]&\footnotemark[2]&\footnotemark[3]\\
 0&10000&9994 &30533 &10000&10771 &47427 &10000&11249 &71936 &10000&9750 &$1.03\times 10^5$ \\
 1&256&258 &137 &250&251 &128 &248&250 &125 &247&247 &122 \\
 2&78&78 &49 &66&66 &37 &64&63 &34 &63&63 &33 \\
 3&46&46 &37 &32&32 &20 &29&29 &17 &28&28 &16 \\
 4&39&39 &39 &19&20 &14 &17&17 &10 &16&16 &9 
\end{tabular}
\end{ruledtabular}
\footnotetext[1]{Statistical average}
\footnotetext[2]{Evaluated with the method of Sec. \ref{continuous}}
\footnotetext[3]{Method of Sec. \ref{continuous}, without fluctuations in $\A_k$}
\end{table*}

Since the disparity between the rows $a$ and $c$ in Table~\ref{bigN} grows as the grid resolution is increased, one should ask whether this effect is due to the necessity of resolving small-scale features, or just due to the lager number of computational cells. In order to answer this question, we repeated our calculations for $N=32$, but increased $\R$, $L/\xi$ and $n$ by a factor of 4. In principle, one should recover the results obtained for $N=8$, and this is indeed found when the fluctuations in $\A_k$ are taken into account. In row $c$, we recover the values for $N=8$ for $n\neq 0$ (i.e., these harmonics behave as ``intensive" quantities), but the value obtained for $n=0$ is roughly multiplied by 4 (behaves as ``extensive").

It might be suspected that our numerical method breaks down for large $N$. This is very unreasonable. First, practically the same results were obtained for different values of $\tau$. Second, if the results without fluctuations in $\A_k$ were numerically wrong, one would need a miracle to correct the results by taking these fluctuations into account.

\subsection{Nonuniform ring\label{non}}
We finally consider a case with nonuniform cross section. We use the gauge-invariant description in Eq.~(\ref{GGLtil}), with $\beta =0$ and consider a ring with just $N=3$ elements. For simplicity, we also take $\xi =L/3$. We take periodic boundary conditions, as in Sec.~\ref{continuous}. We denote the cross sections by $w_1=\mu w_0$ and $w_2=w_3=w_0$. This is an artificial model and clearly we cannot regard its boundary as a smooth surface, but the important feature for the present purpose is that it is a self-consistent model.

The energy can be diagonalized by a linear transformation $\ps_1=\varphi_1+\varphi_2+\varphi_3$, $\ps_{2,3}=\varphi_1-(\mu/2)(\varphi_2+\varphi_3)\pm\nu (\varphi_2-\varphi_3)$, with $\nu=\frac{1}{2}((4+12\mu+7\mu^2+\mu^3)/(7+\mu))^{1/2}$. This transformation gives $G=(\alpha Lw_0/6)(2+\mu)(2|\varphi_1|^2+(2+5\mu+\mu^2)(|\varphi_2|^2+|\varphi_3|^2))$.

As in Sec.~\ref{continuous}, the current becomes a Lagrange multiplier. Since $\Gamma_{A,k}\propto w_k^{-1}$, this time we have to substract a uniform term from every $w_k\A_k$ in order to keep a constant value of $\sum\A_k$.

\begin{figure}
\scalebox{0.85}{\includegraphics{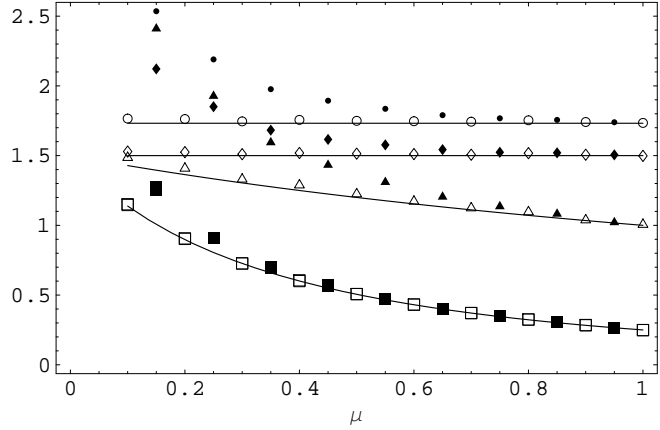}}%
\caption{\label{nonunif} Statistical averages of several fluctuating quantities as functions of uniformity. For a ring with uniform cross section, $\mu =1$. The lines were obtained using statistical mechanics, the empty symbols were evaluated using fluctuations with the sizes dictated by Eqs. (\ref{GamA}) and (\ref{Gamu}) and the filled symbols were evaluated using fluctuations of the same size for the three elements. $\triangle$ $\AL Lw_0\langle |\varphi_1|^2 \rangle $; $\Box$ $\AL Lw_0\langle |\varphi_{2,3}|^2 \rangle $ ($\varphi_{1,2,3}$ as defined in Sec.~\ref{non}); $\diamond$ $\langle G \rangle /2k_BT$; $\circ$ $\langle G^2 \rangle^{1/2} /2k_BT$. Besides $N=3$ and $\xi=\frac{L}{3}$, the parameters are the same as in Fig.~\ref{psin4}.}
\end{figure}

Figure \ref{nonunif} shows the statistical averages of $|\varphi_{1,2,3}|^2$ and of some powers of the energy that are obtained for this model using the method developed in Sec.~\ref{METHOD}. For comparison, we also evaluated these averages using fluctuation sizes that are independent of the local cross section; for the purpose of this comparison we replaced $w_k$ in Eq.~(\ref{GamA}) and Eq.~(\ref{Gamu}) with the geometric average of the cross section, $\mu^{1/3}w_0$.

\subsection{\label{TH}Paraconductivity under constant electric field}
Tucker and Halperin \cite{IofE} evaluated the supercurrent along a thin uniform wire when a longitudinal electric field is applied. The strongest assumption of their model is that the electric field remains constant in time and position, and is not influenced by $\psi$. An unphysical feature of this model is that the total current is not uniform along the wire. The supercurrent is defined by its spatial average, as in Eq.~(\ref{sc1}). An additional assumption was that the length of the wire is infinite. The main effort of Ref.~\cite{IofE} was focused on the influence of the term $\beta |\ps|^4$ in Eq.~(\ref{GGL}), but at this stage we keep $\beta =0$. Their result was
\be
I_{\rm SC}=\frac{ek_BT}{\sqrt{\pi}\hbar}\int_0^\infty dy\, y^{1/2}\exp\left[-\left(\frac{V_0}{V}\right)^{2/3}y-\frac{y^3}{12}\right]\;,
\label{Tucker}
\ee
where $V$ is the applied voltage and $V_0=\alpha L/\gamma e\xi$. $I_{\rm SC}$ is an increasing function of $V$; for $V\gg V_0$, $I_{\rm SC}$ approaches $2ek_BT/\sqrt{3}\hbar =1.15ek_BT/\hbar$.

\begin{figure}
\scalebox{0.85}{\includegraphics{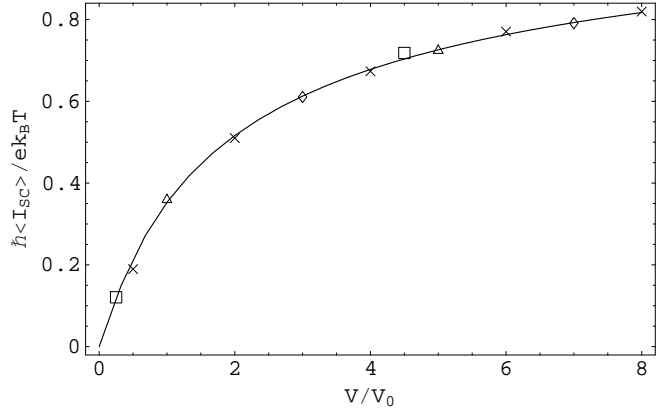}}%
\caption{\label{IE} Average of the superconducting current as a function of the applied voltage under the assumption that the electric field is constant and uniform, in spite of the fluctuations of the superconducting order parameter. The line corresponds to Eq.~(\ref{Tucker}). In all cases $\t=3\times 10^{-5}$, $\tilde\xi =21$ and, unless stated otherwise, $N_{\rm av}=2\times 10^8$, $N_{\rm relax}=6\times 10^7$ and $\AL=10$. $\triangle$ $\xi=L$, $\AL=4$; $\Box$ $\xi=0.3L$; $\diamond$ $\xi=L$; $\times$ $\xi=3L$, $N_{\rm av}=3\times 10^8$, $N_{\rm relax}=9\times 10^7$.}
\end{figure}

In order to reproduce this model, we follow just the evolution of $\ps_k$, whereas $\A_k(t)$ follows from $d\A_k/dt=-2eV/N\hbar $. We take periodic boundary conditions for $\psi$, i.e. $\ps_{N+1}=\exp(-2ieVt/\hbar )\ps_1$, and the condition of infinite length requires $L\gg\xi$. The results are shown in Fig.~\ref{IE} for several cases in the range $0.3\le\xi\le 3L$. There is good agreement with the prediction of Ref.~\cite{IofE} but, surprisingly, the requirement of large $L$ seems to be unnecessary.

\section{APPLICATION---PARACONDUCTIVITY\label{APPL}}

\subsection{Voltage--current characteristic of a uniform wire above the critical temperature\label{vol-cur}}

\subsubsection{Fixed voltage\label{vol-cur1}}
We are now in a position that enables us to evaluate self-consistently the average current $\langle I\rangle$ along a superconducting wire when a fixed voltage $V$ is applied. The procedure is similar to that of Sec.~\ref{TH}, but this time the evolution of $\A_k$ is given by Eq.~(\ref{algo}). The instantaneous total current $I$ has to be such that in every step $\sum\A_k$ decreases by $2eV\tau/\hbar$. $\langle I\rangle$ is obtained as the average of $I$ over $N_{\rm av}$ iteration steps.

The dashing lines in Fig.~\ref{VI} show the values obtained for $\langle I\rangle(V)$ in the range $10^{-1}\alt I\hbar /ek_BT\alt 10^{3/2}$ and $40\le\R\le 25000$. For $\R=40$ these values are scarcely below the currents that would be present in a normal material (in a logarithmic scale); this means that supercurrents are small compared to the normal currents. For larger values of $\R$, Ohm's law is approached in the large current extreme, whereas in the low current extreme $\langle I\rangle$ is not proportional to $V$. It might be argued that $\langle I\rangle$ could become proportional to $V$ for still lower currents; we did not attempt to evaluate $\langle I\rangle$ for lower currents, because Eq.~(\ref{44}) predicts fluctuations of $I_{\rm SC}\hbar /ek_BT$ of the order of $\sqrt{\xi/L}$ for a single step and, since the number of independent values of $I_{\rm SC}$ is of the order of $N_{\rm av}\t\sim 10^3$, we should at best expect $\langle I\rangle\hbar /ek_BT$ to be accurate to the order of $10^{-2}$.

\subsubsection{Fixed current}
Another experimental situation is prescribed by keeping a fixed current $I$. In this case we are interested in the average value of the voltage. The evolution of $\ps$ and $\A$ is the same as in Sec.~\ref{continuous}, except for an additional term $\R\t\hbar I/2N\AL ek_BT$ in the increment of each $\A_k$ in every step. The boundary conditions are dictated by the experimental setup; in the absence of experimental details
we shall just assume that the coupling between $\psi_1$ and $\psi_0$ (and between $\psi_N$ and $\psi_{N+1}$) can be neglected. In order to avoid a significant influence of the boundary conditions at the ``contacts," we adopted a ``four-probe technique," i.e., the voltage was measured between the segments $1+N_{\rm cont}$ and $N-N_{\rm cont}$ and the result was multiplied by $N/(N-2N_{\rm cont})$. 

 The average voltage drop $\langle V\rangle$ is given by
\be
\frac{\gamma e\langle V\rangle}{k_BT}=-\frac{N}{N-2N_{\rm cont}}\,\frac{\AL}{2N_{\rm av}\t}\sum \Delta \A_k \;,
\label{VofI}
\ee
where $\Delta \A_k$ is the increment of $\A_k$ in a step and the sum is over the segments $1+N_{\rm cont}\leq k\leq N-N_{\rm cont}$ and also over the $N_{\rm av}$ iteration steps.

\begin{figure}
\scalebox{0.85}{\includegraphics{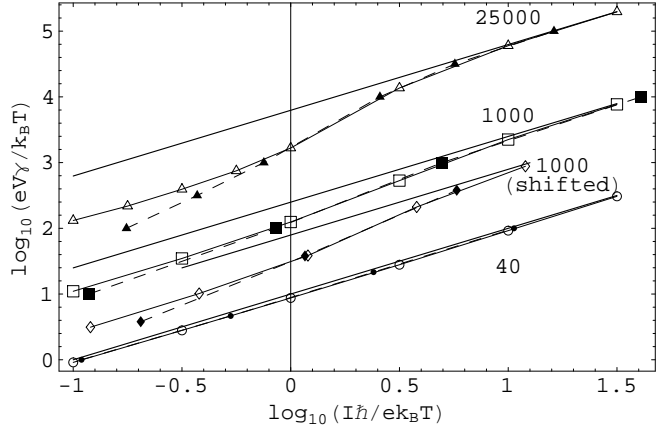}}%
\caption{\label{VI}Voltage--current characteristic. Filled symbols (dashing lines) stand for $\langle I\rangle (V)$ and empty symbols (continuous lines) for $\langle V\rangle (I)$. The straight lines correspond to Ohm's law in the normal state. Unless stated otherwise, $N=21$, $N_{\rm cont}=3$, $N_{\rm av}=2\times 10^8$, $N_{\rm relax}=6\times 10^7$, $\AL=10$, $\xi=L/\sqrt{10}$ and $\t=10^{-5}$. The values of $\R$ are marked next to the curves in the large-current regime. The lines with the rhombs correspond to $\R=1000$ and $\AL=3$; for visibility, they have been lowered by half a unit. For $\R=25000$, several points of $\langle I\rangle (V)$ were evaluated using $\t=3\times 10^{-5}$ and several points of $\langle V\rangle (I)$ were evaluated using $\t=3\times 10^{-6}$, $N_{\rm av}=10^8$, $N_{\rm relax}=3\times 10^7$; these changes of computational parameters do not produce noticeable changes in this figure.}
\end{figure}

The continuous lines in Fig.~\ref{VI} show our results for $\langle V\rangle (I)$. For large and intermediate currents $\langle V\rangle (I)$ coincides with $\langle I\rangle(V)$ in the scale of the graph, but for low currents we see that they are different curves. This difference reflects the fact that a situation in which $I$ is kept fixed and $V$ fluctuates is not equivalent to a situation in which $V$ is kept fixed and $I$ fluctuates. This time we do reach a small current regime where $\langle V\rangle$ is proportional to $I$, with an effective resistance that is smaller than $L/w_0\sigma$.

The difference between $\langle V\rangle (I)$ and $\langle I\rangle(V)$ has been dramatically observed in recent experiments \cite{S-shape,MichotteB69}.

\subsection{Low current limit\label{Low}}
Let us denote by $\sigma '$ the contribution of the supercurrent to the electric conductivity. Taking the limit $V\rightarrow 0$ in Eq.~(\ref{Tucker}) we obtain
\be
\sigma '=\sigma\R\xi/8\AL L \;.
\label{preAL}
\ee
This result was initially obtained by Aslamazov and Larkin \cite{asla} and the value of $\sigma '$ according to Eq.~(\ref{preAL}) will be denoted by $\sal$.  It can also be written as
\be
\sal=\R\sigma\L^2(\xi/2L)^3=\R\sigma /8\AL^{3/2}\L =K\ep^{-3/2} \;,
\label{AL}
\ee
where $\ep=(T-T_c)/T_c$, $T_c$ is the critical temperature and $K=\pi e^2\xi (0)/16\hbar w_0$. The last form of Eq.~(\ref{AL}) neglects terms of order $\ep^{-1/2}$ and assumes $\alpha =8k_B(T-T_c)\gamma /\pi$.

We will now evaluate $\sigma '$ taking the electromagnetic fluctuations into account. This will be done by fixing a small current $I$, evaluating the average voltage $\langle V\rangle$ as discussed in the previous section, and then $\sigma '$ will be given by
\be
\frac{\sigma '}{\sigma}=\frac{\R \hbar I/ek_BT}{4\gamma e\langle V\rangle /k_BT}-1 \;.
\label{oursig}
\ee
The choice of $I$ as the nonfluctuating quantity is based both on Fig.~\ref{VI} and on the fact that most experiments are performed under this condition.

The current has to be sufficiently small in order to yield a voltage proportional to it. Our requirement was $I_{\rm SC}\alt 0.1ek_BT/\hbar$. This value was inspired by Figs. \ref{IE} and \ref{VI} and is in agreement with experiments \cite{TH,Giordano,early}, that required $I\alt 3ek_BT/\hbar$. For experiments that used larger currents, the conductivity was current dependent in the interesting range \cite{cook}. Our requirement is translated into the condition $I\alt 0.1(1+\sigma /\sigma ')ek_BT/\hbar$. Since $\sigma '$ was not known in advance, we used $I=0.1(1+\sigma /\sal)ek_BT/\hbar$, which turned out to be appropriate in all cases. In several cases we doubled $I$ and verified that $\langle V\rangle$ was indeed doubled.

\begin{figure}
\scalebox{0.85}{\includegraphics{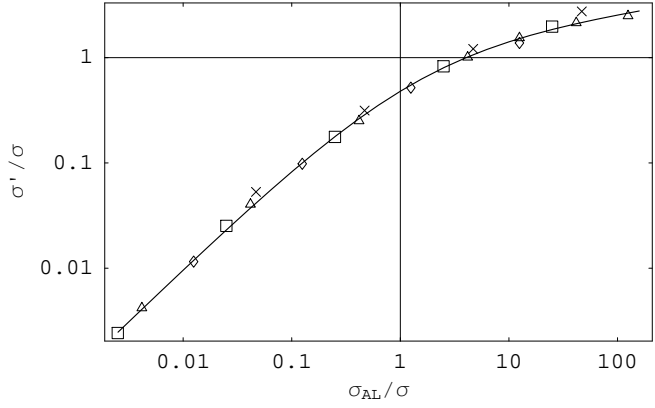}}%
\caption{\label{ourAL} Superconducting contribution to the electric conductivity, evaluated as described in Sec.~\ref{Low} and displayed as a function of the contribution that would be obtained according to Eq.~(\ref{preAL}). The line is an empiric function. In all cases $N_{\rm av}=2\times 10^8$ and $N_{\rm relax}=6\times 10^7$. For $\sal<10\sigma$, $\t=0.003\AL/\R$; otherwise, $\t=0.03\AL/\R$.  $\triangle$ $\R=5000$, $\xi=0.2L$, $N=14$, $N_{\rm cont}=2$; $\Box$ $\R=20$, $\xi=0.2L$, $N=14$, $N_{\rm cont}=2$; $\diamond$ $\R=100$, $\xi=0.1L$, $N=28$, $N_{\rm cont}=4$; $\times$ $\R=5000$, $\xi=0.3L$, $N=14$, $N_{\rm cont}=3$.}
\end{figure}

Figure \ref{ourAL} shows the results obtained for $\sigma '$, for several values of $\R$ and $\xi/L$, while $\AL$ is varied over typically four orders of magnitude. Most of the results sit on a universal curve. The line in the graph is 
\be
\sigma '/\sigma =((\sal/\sigma)^{-4/5}+(4/5)(\sal/\sigma)^{-1/8})^{-5/4} 
\label{empirAL}
\ee
and seems to fit the results reasonably well in this range. The results for $\xi=0.3L$ are too high. We suspect that for this large $\xi/L$ ratio the conductivity is influenced by the boundary conditions; it is known that when $\xi$ is of the order of the length of the wire, qualitatively different behavior is obtained \cite{anomaly}. 

Far from the critical temperature Eq.~(\ref{empirAL}) coincides with the Aslamazov--Larkin result, but close to the critical temperature the conductivity increases at a slower rate, even in the case $\beta =0$. 

\subsection{Inclusion of the nonlinear term \label{beta}}
Until now we have neglected the term with higher order in $\psi_k$. This can be done for temperatures that are sufficiently above the critical temperature. We are now interested in the region close or below the critical temperature and we therefore set $\beta \neq 0$. Except for the general treatment in Sec.~\ref{Num}, this section will consider a wire with uniform cross section $w_0$. In the general case, $w_0$ stands for a typical cross section.

\subsubsection{Normalized evolution equations\label{Num}} We start by defining some useful quantities. $\bar\psi$ and $\xi_\beta$ are defined by requiring $\beta\bar\psi^4 w_0\xi_\beta =k_BT$ and $\beta\bar\psi^2 \xi_\beta^2 =\hbar^2/2m$. The implication of these requirements is that $\bar\psi$ will be an estimate of the size of the order parameter at the critical temperature, and $\xi_\beta$ will be an estimate of the coherence length at this temperature. Solving these two equations gives
\begin{eqnarray}
\bar\psi&=&(2mk_B^2T^2/\beta \hbar^2w_0^2)^{1/6} \;, \label{psibar}\\
\xi_\beta&=&(\hbar^4w_0/4\beta m^2k_BT)^{1/3} \;. \label{xibeta}
\end{eqnarray}
In terms of more directly measurable quantities we can write 
\be
\xi_\beta=(w_0\Phi_0^2/32\pi^3\kappa^2k_BT)^{1/3} \;, 
\label{betakap}
\ee
where $\kappa$ is the Ginzburg--Landau parameter; for a wire of cross section $w_0\sim 10^{-11}\,{\rm cm}^2$, critical temperature $T\sim 1\,$K and $\kappa\sim 0.1$, $\xi_\beta\sim 7\times 10^{-4}\,$cm. We also define the normalized order parameter $\psi_{\beta k}=\psi_k/\bar\psi$ (and similarly, $\ps_{\beta k}=\ps_k/\bar\psi$), the normalized step size $\t_\beta =\tau \beta\bar\psi^2/\gamma \hbar $ and the ratio $\alpha '=\alpha /\beta\bar\psi^2=\AL(\xi_\beta\L/L)^2$; if $\alpha$ is a perfectly linear function of the temperature, then $\alpha '=(\xi_\beta/\xi (0))^2\epsilon$.

With these notations, the ``macroscopic" increments of $\ps_{\beta k}$ and $\A_k$ during a time step $\tau$ are
\begin{eqnarray}
\Delta_{\rm mac}\ps_{\beta k}&=&-\t_\beta [(\alpha '+|\ps_{\beta k}|^2)\ps_{\beta k}+
(N\xi_\beta/L)^2 (2w_k)^{-1} \nonumber \\
&&[(w_k+w_{k+1})(\ps_{\beta k}-\ps_{\beta k+1})+(w_k+w_{k-1})(\ps_{\beta k}-\ps_{\beta k-1})]]\;, \label{Dpsibeta}\\
\Delta_{\rm mac}\A_k&=&-(\t_\beta \R/2w_k)(\xi_\beta\L/L)^2[(w_0/N)(I\hbar /ek_BT)+\nonumber \\
&&(\xi_\beta/L){\rm Im}[(w_{k-1}+w_{k})\ps_{\beta k-1}^*\ps_{\beta k}+(w_{k}+w_{k+1})\ps_{\beta k}^*\ps_{\beta k+1}]] \;. \label{DAbeta}
\end{eqnarray}
The numerical details for the integration of these equations are discussed in Appendix \ref{AA}.

The fluctuations in the increment of $\A_k$ have a variance $2\R\t_\beta (w_0/Nw_k)(\xi_\beta\L/L)^2$; the fluctuations of $\ps_{\beta k}$ are taken into account by adding to the real and to the imaginary part terms with variance $N\t_\beta w_0\xi_\beta /w_k L$ and then modifying the phase according to Eq.~(\ref{algotil}).

\subsubsection{\label{results}Basic results for low currents} 
\begin{figure}
\scalebox{0.85}{\includegraphics{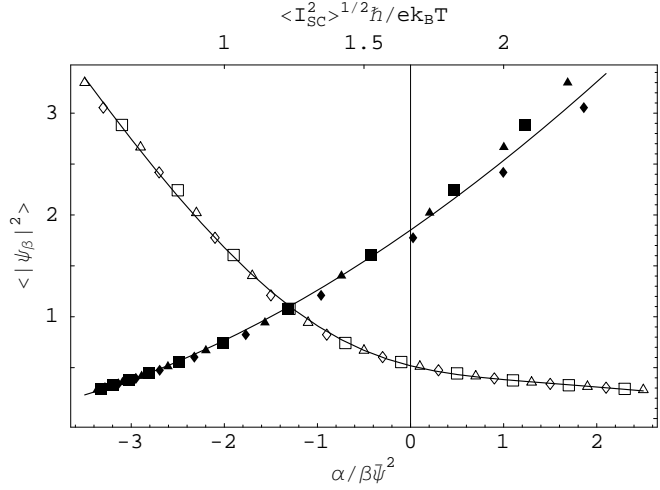}}%
\caption{\label{psisq}Average amount of superconducting pairs as a function of $\alpha $, close to the critical temperature. Empty symbols stand for $\langle |\psi_{\beta}|^2\rangle (\alpha ')$ (lower abscissa) and are fitted by Eq.~(\ref{empirpsi});
filled symbols stand for $\langle |\psi_{\beta}|^2\rangle (\langle I_{\rm SC}^2\rangle)$ (upper abscissa) and are fitted by
$\langle I_{\rm SC}^2 \rangle ^{1/2}=2.61(\xi_\beta /L)^{1/2}\langle |\psi_\beta |^2\rangle^{0.58}ek_BT/\hbar $.
$N=21$, $N_{\rm cont}=2$, $N_{\rm av}=4\times 10^7$, $N_{\rm relax}=1.2\times 10^7$, $\xi_\beta=0.2L$, $\L=10$, $\t_\beta =5\times 10^{-4}$. $\triangle$ $\R=0.1$; $\Box$ $\R=3$; $\Diamond$ $\R=100$.}
\end{figure}

Figure \ref{psisq} shows the average value of $|\psi_{\beta k}|^2$, obtained using Eqs.~(\ref{Dpsibeta})-(\ref{DAbeta}) and Eq.~(\ref{algointern}) for a wire of uniform cross section and in a region that includes the critical temperature, as $\AL$ and $\R$ were varied. $|\psi_{\beta k}|^2$ was averaged over $N_{\rm av}$ iteration steps and also over $k$ in the range $1+N_{\rm cont}\le k\le N-N_{\rm cont}$. In several cases $\AL$ was swept downwards and then upwards, in order to check for hysteresis; for the parameters we studied, no evidence for hysteresis was found. As could be expected from Eq.~(\ref{Dpsibeta}), the values of $\langle |\psi_{\beta}|^2\rangle$ are a universal function of $\alpha '$. We also evaluated $\langle |\psi_{\beta}|^2\rangle$ for different values of $\xi_\beta /L$ and $\L$ (not shown in Fig.~\ref{psisq}), and the results lie on the same curve. 

In the considered range, $\langle |\psi_{\beta}|^2\rangle$ can be fitted by the empiric expression
\be
\langle |\psi_{\beta}|^2\rangle=-1.44(\alpha '/2-\log_e(1+e^{\alpha '/2}))-0.70(\alpha '-\log_e(1+e^{\alpha '}))-1.93/(1+e^{\alpha '}) \;.
\label{empirpsi}
\ee
As in \cite{space}, our results have a hardly visible inflection point near $\alpha '\sim -2.5$; our fit (\ref{empirpsi}) is not sufficiently accurate to reproduce this feature. As a test for linear response, we evaluated $\langle |\psi_\beta |^2\rangle$ in several cases for zero current and for $I=0.1ek_BT/\hbar$; the current caused a decrease of $\langle |\psi_\beta |^2\rangle$ of the order of 0.2\%.

We also evaluated $\langle I_{\rm SC}^2 \rangle$ for zero total current and various values of $\alpha '$, $\xi_\beta /L$, $\L$ and $\R$. Some of the results are shown in Fig.~\ref{psisq}. 

\begin{figure}
\scalebox{0.85}{\includegraphics{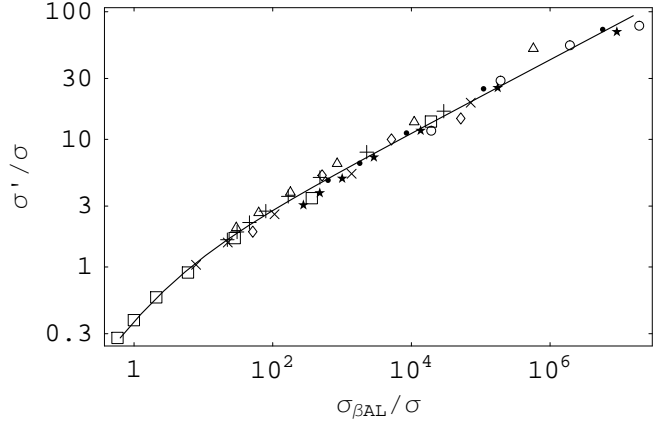}}%
\caption{\label{sigmabeta} Superconducting contribution to the electric conductivity in a region where the order parameter is not small. The symbols were evaluated using Eq.~(\ref{oursig}). The line was evaluated using Eq.~(\ref{modifiedAL}). Unless stated otherwise, $N=21$, $N_{\rm cont}=N/7$, $N_{\rm av}=2\times 10^8$, $N_{\rm relax}=6\times 10^7$, $\xi_\beta=0.2L$ and $\L=100$. $\t_\beta =10^{-4}$ for $\R <30$ and $\L <400$; otherwise, $\t_\beta =10^{-5}$. $I=0.1ek_BT/\hbar$, except for the two leftmost squares, for which $I=0.2ek_BT/\hbar$. $\Box$ $\R=0.1$; $\triangle$ $\R=3$; $\bullet$ $\R=30$;$+$ $\R=2$, $\L=200$; $\star$ $\R=3$, $\L=400$;  $\times$ $\R=3$, $\xi_\beta=0.1L$, $N=42$;  $\diamond$ $\alpha '=-1.146$ ($\langle |\psi_\beta |^2\rangle=1$); $\circ$ $\alpha '=-2$.}
\end{figure}

Figure \ref{sigmabeta} shows the superconducting contribution to the electric conductivity for several values of $\R$, $\L$ and $\xi_\beta$, with $\alpha '$ in the range $-2\le\alpha '\le 2$. The results were fitted by a modification of Eq.~(\ref{empirAL}). Writing $K'=\R\xi_\beta^3\L^2/L^3$, we defined the quantity
\be
\sigma_{\beta\rm{AL}}=(K'\sigma /8)(\alpha '+\langle |\psi_\beta |^2\rangle +0.3488\tanh (0.734 \langle |\psi_\beta |^2\rangle^{1.5}))^{-3/2}\;,
\label{sbal}
\ee
which is a generalization of $\sal$ and reduces to it in the limit $\psi_\beta\rightarrow 0$. $\langle |\psi_\beta |^2\rangle$ was evaluated using Eq.~(\ref{empirpsi}). Finally, the fit used for $\sigma '$ was
\be
\sigma '/\sigma =((\sigma_{\beta\rm{AL}}/\sigma)^{-4/5}+1.205(\sigma_{\beta\rm{AL}}/\sigma)^{-0.23})^{-5/4} \;.
\label{modifiedAL}
\ee
 
\subsubsection{Voltage distribution \label{VD}}
In the previous sections we have evaluated the time average of the voltage. In this section we investigate how the voltage fluctuates over time steps. We define $L'=(1-2N_{\rm cont}/N)L$, the length of the wire excluding the contact regions, and the dimensionless normalized voltage $\tilde V=\gamma eV/\beta\bar\psi^2$. For a single step, the normalized voltage drop over $L'$ is $\tilde V=-\sum_{k=1+N_{\rm cont}}^{N-N_{\rm cont}} \Delta \A_k/2\t_\beta$.

\begin{figure}
\scalebox{0.85}{\includegraphics{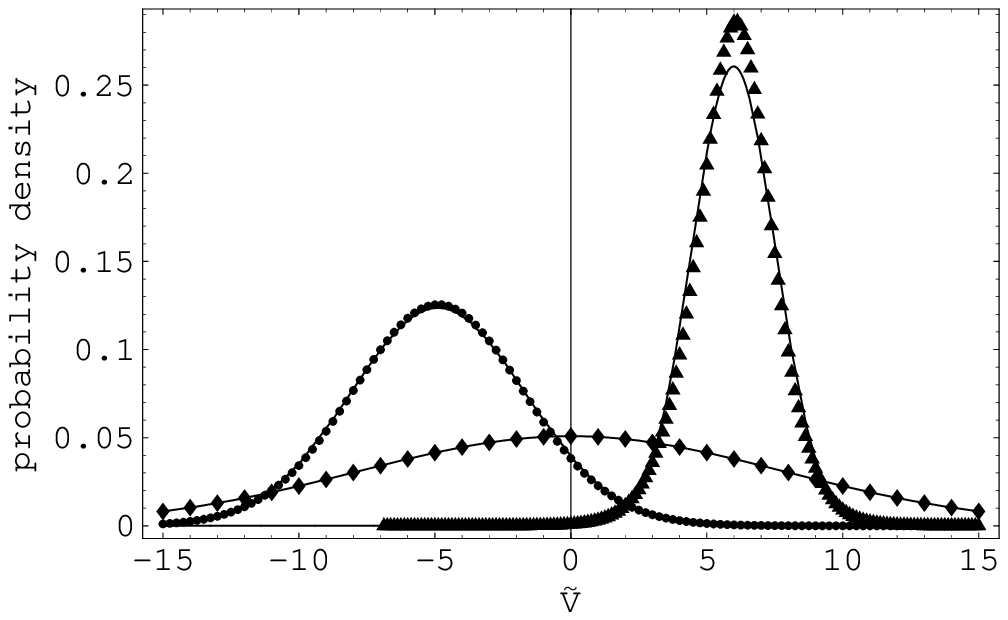}}%
\caption{\label{histog} Distribution of the voltage drop along the wire, evaluated for single steps. $L'=3.2\xi_\beta$, $I=ek_BT/\hbar$, $\L=10$, $\R=1$, $N=30$, $N_{\rm cont}=3$, $\t_\beta =5\times 10^{-4}$, $N_{\rm av}=3\times 10^8$, $N_{\rm relax}=6\times 10^7$.
The solid line is the gaussian distribution 
$(2\pi(\langle\tilde V^2\rangle-\langle\tilde V\rangle^2))^{-1/2}\exp[-(\tilde V-\langle\tilde V\rangle)^2/(2(\langle\tilde V^2\rangle-\langle\tilde V\rangle^2))]$.
$\triangle$ $\alpha '=-0.4$; $\bullet$ $\alpha '=-3$; $\diamond$ $\alpha '=-16$. For visibility, the line and symbols for $\alpha '=-0.4$ (respectively, $-3$) have been shifted 5 units to the right (left).}
\end{figure}

Figure \ref{histog} is a set of histograms for the values of $\tilde V$ obtained in every step, for several values of $\alpha '$. $\tilde V$ has contributions from $\Delta_{\rm mac}\A_k$ in Eq.~(\ref{DAbeta}) and also from the fluctuations of $\A_k$. Since the fluctuations of $\A_k$ are known to be gaussian, the results we present take into account the contribution from $\Delta_{\rm mac}\A_k$ only. These histograms have been fitted by gaussian distributions.

\begin{figure}
\scalebox{0.85}{\includegraphics{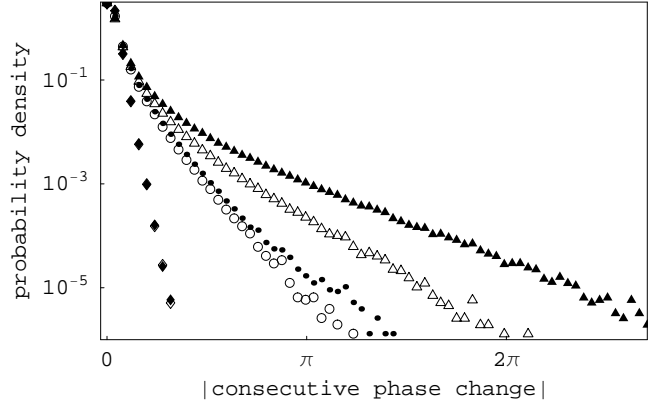}}%
\caption{\label{consecutive} Distribution of the absolute values of ``consecutive phase changes." Empty symbols are for positive changes of the phase and filled symbols for negative changes. The parameters, including symbol identity, are the same as in Fig.~\ref{histog}.}
\end{figure}

From Fig.~\ref{histog} we might conclude that the voltage distribution is approximately gaussian, but such a conclusion would be misleading, since the voltage drops for steps that are close in time are correlated. A more informative distribution is that of the ``consecutive phase change," which we define as follows. For every step we check the sign of the macroscopic contribution to the phase increment, $\sum_{k=1+N_{\rm cont}}^{N-N_{\rm cont}} \Delta \A_k$, and add all the consecutive increments until the sign is reversed. Note that in order to decide when to stop adding increments and start evaluating the next phase change we monitored only the macroscopic contribution to $\sum\Delta\A_k$, but in order to evaluate these increments we took into account the microscopic fluctuations as well. The results are shown in Fig.~\ref{consecutive}. Within statistical accuracy, all the curves decay monotonically. Although the sizes of typical voltage drops in a single step increase when $\alpha '$ becomes more negative, we see that the probability of large phase excursions decreases with $|\alpha '|$. We also see that the probability of large negative phase changes is greater than that of large positive phase changes, and this explains the average voltage drop.

According to the Langer--Ambegaokar--McCumber--Halperin (LAMH) scenario \cite{LAMH}, the voltage drop is due to phase slips, in which the phase decreases by $2\pi$. Accordingly, we expected to find a relatively large probability density near $2\pi$ in Fig.~\ref{consecutive}, but we did not obtain support for this view. It might seem that this negative result contradicts positive evidences, such as the observations by Lukens and Goodkind \cite{LG} or the simulations in \cite{Tarlie}, but these evidences correspond to a closed ring setup, and not to a fixed current. It has been further claimed that for a long wire the existence of phase slips ought to be insensitive to the boundary conditions (page 1088 in \cite{ST}), but this is not the situation studied here.
Phase slips will be reconsidered in Sec.~\ref{High}. 

\subsubsection{Scaling with length}
In the previous sections we have tacitly assumed that the voltage drop along a uniform wire is proportional to its length. We have taken advantage of this assumption and considered short wires, that do not require large computational resources. In this section we check this assumption. We have evaluated the voltage and its variance for several values of $\alpha '$ and several lengths, while intrinsic quantities such as $\R/\L$ are kept at a fixed value. The results are summarized in Table~\ref{long}. 
The average of the voltage per unit length, and its variance per unit length, which should in principle be independent of $L'$, seem to increase slightly with $L'$; in the case of $\langle V\rangle /L'$, this trend seems to saturate within the considered range. For the parametrers in Table~\ref{long} the deviations from the proportionality $\langle V\rangle\propto L'$ are most likely due to the finite sizes of the elements of the computational grid and of the contact regions; on the other hand, this proportionality breaks down completely when additional length brings about an additional phase-slip center \cite{oldexp}.

\begin{table*}
\caption{\label{long}Voltage drop and its variance, per unit length of the wire. $L'$ is the length (excluding the contact regions) over which $V$ is evaluated. The parameters used are $\R/\L=0.033$, $I=0.3ek_BT/\hbar$, $\xi_\beta =7.2\hbar/\sqrt{2mk_BT}$, $N_{\rm cont}=3$, $N=6+3.6L'/\xi_\beta$, $\t_\beta =5\times 10^{-4}$, $N_{\rm av}=2\times 10^8$, $N_{\rm relax}=6\times 10^7$.}
\begin{ruledtabular}
\begin{tabular}{c|ccc|ccc}
 &\multicolumn{3}{c|}{$10^3(\gamma e/k_BT)\xi_\beta\langle V\rangle /L'$}&\multicolumn{3}{c}
{$10^2(\gamma e/k_BT)^2\xi_\beta(\langle V^2\rangle -\langle V\rangle^2)/L'$}\\ \hline
 $\alpha '$&$L'=2.5\xi_\beta$&$L'=10\xi_\beta$&$L'=30\xi_\beta$&$L'=2.5\xi_\beta$&$L'=10\xi_\beta$&$L'=30\xi_\beta$ \\
 -0.5& 8.48&8.87&8.97&1.52&1.58&1.75\\
 -1&5.89&6.31&6.42&2.11&2.15&2.23\\
 -2&1.26&1.31&1.30&4.35&4.48&4.52 
\end{tabular}
\end{ruledtabular}
\end{table*}

\subsubsection{High-current phase slips \label{High}}
The phase slips in the LAHM scenario are assumed to be caused by thermal fluctuations. We unsuccessfully looked for them in Sec.~\ref{VD}. On the other hand, early experiments \cite{oldexp,Tidecks} show steps in the voltage-current characteristic, that have been attributed to phase slips. These steps appear at currents that are much larger than the current that characterizes the scale for thermal fluctuations, $ek_BT/\hbar$. Modern versions of these experiments are the subject of recent investigations \cite{S-shape,MichotteB69}. A numeric study using the same simple TDGL model as in the present paper was performed by Kramer and Baratoff \cite{Barat}.

According to the basic phase-slip idea, there are moments at which $\psi$ vanishes at places called phase-slip centers. When this happens, the order parameter releases part of its winding by changing by $2\pi$ the phase difference across this center. This phase change is equivalent to a change in the electromagnetic potential $A$, and gives rise to a voltage pulse. According to typical models, phase-slip centers are located at fixed positions.

\begin{figure}
\scalebox{0.85}{\includegraphics{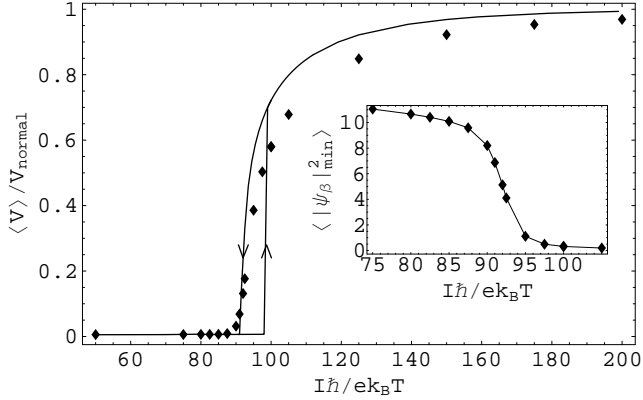}}%
\caption{\label{vnormal}Voltage-current characteristic for high currents. $V_{\rm normal}$ stands for $L'I/\sigma w_0$ and $\langle\cdots\rangle$ is the average over $N_{\rm av}$ steps. The line (with the hysteresis loop) is obtained when fluctuations are ignored and the symbols take fluctuations into account. The inset shows the minimum value of the order parameter as a function of the current. $\alpha '=-16$, $L'=3.2\xi_\beta$, $\L=10$, $\R=1$, $N=30$, $N_{\rm cont}=3$, $\t_\beta =5\times 10^{-4}$, $N_{\rm av}=N_{\rm relax}=6\times 10^7$.}
\end{figure}

Figure~\ref{vnormal} shows the ratio between the average voltage and the voltage that would be obtained for the same current if the wire were in the normal state. The continuous line was obtained when thermal fluctuations were ignored, i.e. $\eta_{u,v}$ and $\eta_A$ were set equal to zero. In this case we find hysteresis. The symbols were obtained using the actual values of $\eta_{u,v,A}$ required by the fluctuation-dissipation theorem. For the parameters used, the inclusion of fluctuations leads to results that are independent of history. 

We denote by $|\psi|_{\rm min}$ (and similarly $|\psi_\beta |_{\rm min}$) the minimum value of $|\psi_k|$ among the elements $N_{\rm cont}+1\le k\le N-N_{\rm cont}$ for a given time step. The inset in Fig.~\ref{vnormal} shows, as expected, that the region of fast increase of the average voltage is the same as the region of fast decrease of $\langle |\psi_\beta |_{\rm min}^2\rangle$. In the absence of fluctuations, the smallest values of $|\psi_k|$ appear always in the middle of the wire, but, when fluctuations are included, the positions of the minima are widely spread.

\begin{figure}
\scalebox{0.85}{\includegraphics{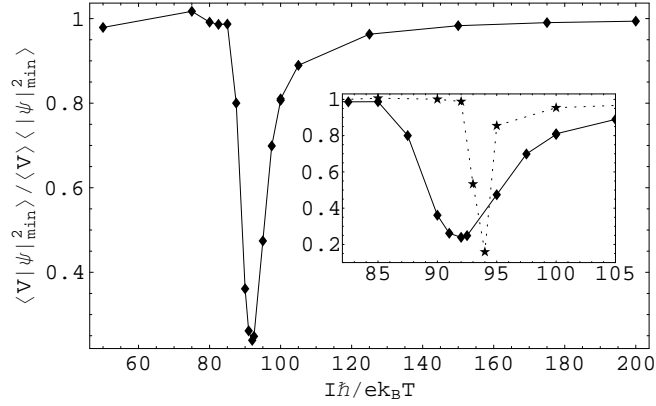}}%
\caption{\label{pV} The ratio $\langle V|\psi|_{\rm min}^2\rangle /\langle V\rangle \langle |\psi|_{\rm min}^2\rangle$ measures the correlation between the voltage and the minimum size of the order parameter. This ratio has to be less than 1 if phase slips contribute to the voltage. The inset shows the region where correlation is significant. The parameters are the same as in Fig.~\ref{vnormal}. $\diamond$ Fluctuations properly taken into account; $\star$ size of fluctuations divided by 2.}
\end{figure}

If phase slips give a considerable contribution to the average voltage, then we expect that the voltage will be large for steps that have small $|\psi|_{\rm min}$, since ideally $|\psi|_{\rm min}=0$ when a phase slip occurs. It follows that $V$ and $|\psi|_{\rm min}^2$ will be negatively correlated and $\langle V|\psi|_{\rm min}^2\rangle /\langle V\rangle \langle |\psi|_{\rm min}^2\rangle$ will be less than 1. Figure~\ref{pV} shows this ratio as a function of the current and we see that there is indeed a region where $\langle V|\psi|_{\rm min}^2\rangle /\langle V\rangle \langle |\psi|_{\rm min}^2\rangle <1$. We expect that this is a region where phase slips occur. The position of the minimum of $\langle V|\psi|_{\rm min}^2\rangle /\langle V\rangle \langle |\psi|_{\rm min}^2\rangle$ nearly coincides with the current at the lower vertex in the hysteresis loop shown in Fig.~\ref{vnormal}, but the dip extends to currents beyond the hysteresis region. The inset in Fig.~\ref{pV} is a close-up at the dip region. In order to appreciate the influence of fluctuations, $\langle V|\psi|_{\rm min}^2\rangle /\langle V\rangle \langle |\psi|_{\rm min}^2\rangle$ was also evaluated with $\eta_{u,v,A}$ set to half their actual values. We see that fluctuations extend the dip region in both directions, but mainly towards lower currents.

\begin{figure}
\scalebox{0.85}{\includegraphics{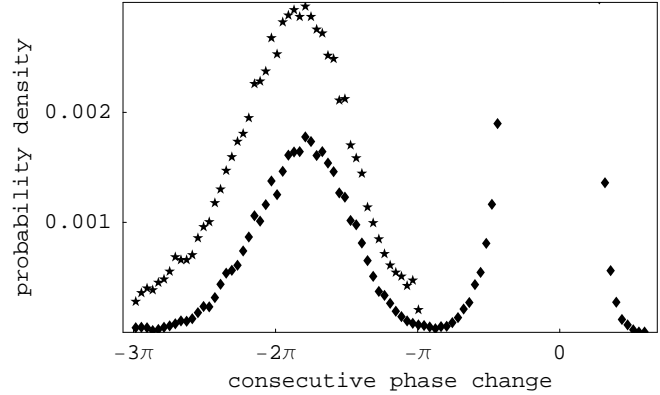}}%
\caption{\label{slippi}Distribution of consecutive phase changes for $I\hbar /ek_BT=92$. The parameters are the same as in Fig.~\ref{vnormal}. The peak near zero has been chopped in order to make phase slips visible. $\diamond$ Phase changes restricted to the region above $-3\pi$; $\star$ phase changes below $-3\pi$ were brought into the region between $-3\pi$ and $-\pi$ by addition of the appropriate integer multiple of $2\pi$.}
\end{figure}

Figure~\ref{slippi} shows the distribution of consecutive phase changes, as defined in Sec.~\ref{VD}, for $I\hbar /ek_BT=92$. We expect from Figs. \ref{vnormal} and \ref{pV} that for this current phase slips will be present. Most of the consecutive phase changes are close to zero, and have been chopped in the figure. However, there is also a significant amount of cases for which the consecutive phase change is close to $-2\pi$, as predicted.

For this large current, the sign of the phase change is not necessarily reversed between phase slips. Therefore, the consecutive phase change may involve more than one phase slip and the consecutive phase change should ideally equal an integer multiple of $-2\pi$. In order to take this possibility into account, we added integer multiples of $2\pi$ to phase changes that were smaller than $-3\pi$, until we brought them to the region between $-3\pi$ and $-\pi$. The results obtained in this way are shown by the stars in Fig.~\ref{slippi}.

\subsubsection{Comparison with experiments}
It is usually accepted that different models are required for the description of the electric conductivity in several temperature ranges in the vicinity of the critical temperature. Above the critical temperature, conductivity is believed to be dominated by the exchange of ``cooperons" between normal electrons \cite{Maki}. Very near the critical temperature, a promising method is the Hartree-Fock approximation to TDGL \cite{IofE}. Below the critical temperature, but not too near it, the LAMH model gives good fits \cite{LAMH}. For still lower temperatures there is a ``foot" which in some cases has been attributed to the contacts \cite{Tinkham} and in others to macroscopic quantum tunneling \cite{ph-slip,Giordano}.
Moreover, the parameters we have studied in section \ref{results} differ by several orders of magnitude from those of the available experimental data.
In spite of this situation, it is instructive to check how well our naive formalism can fit the experimental data.

We analyzed samples of aluminum \cite{TH}, indium \cite{Giordano} and tin \cite{early}, in a temperature range where $10^{-2}\sigma\alt\sigma '\alt10^{1.5}\sigma$. We chose the samples so that they could be considered as one-dimensional, sufficient data in the analyzed range were available and all the required properties of the sample (e.g. length) were reported. We denote by $T'_c$ the temperature that the experimenters considered to be the critical temperature. The data for aluminum are above $T'_c$ and the others below $T'_c$. The normal resistance of the tin sample was smaller by more than three orders of magnitude than those of the other samples; this was due both to its larger cross section and to its longer mean free path. The current passed through the tin sample was slightly above the linear range.

The properties of a sample were estimated as follows. The length was measured by means of a microscope; the cross section was either estimated from microscopic observation or from the critical current. Assuming the geometry is known, the mean free path can be obtained from the normal conductivity \cite{length}. The BCS coherence length and the London penetration depth are assumed to be the same as in bulk material, in spite of the fact that the critical temperature is significantly different than that in bulk. $\xi(0)$ and $\kappa$ can be obtained from these values and the mean free path \cite{Tinkham}, and $\xi_\beta$ from Eq.~(\ref{betakap}). $T'_c$ is either assumed equal to the critical temperature of a reference sample with large cross section, or is taken from a fit to some model. For $\gamma$ we take the value $\pi\hbar^2/16k_BT'_cm\xi(0)^2$.

\begin{figure}
\scalebox{0.85}{\includegraphics{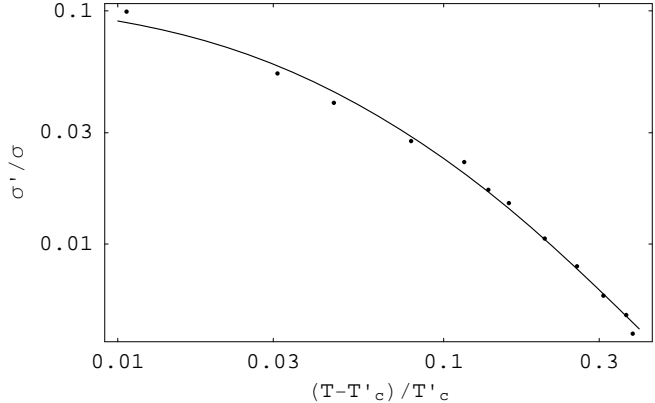}}%
\caption{\label{leib} Paraconductivity of an aluminum microwire as a function of the temperature above $T'_c$. The dots are experimental data and the curve is a fit using Eqs. (\ref{empirpsi})--(\ref{modifiedAL}).}
\end{figure}

\begin{figure}
\scalebox{0.85}{\includegraphics{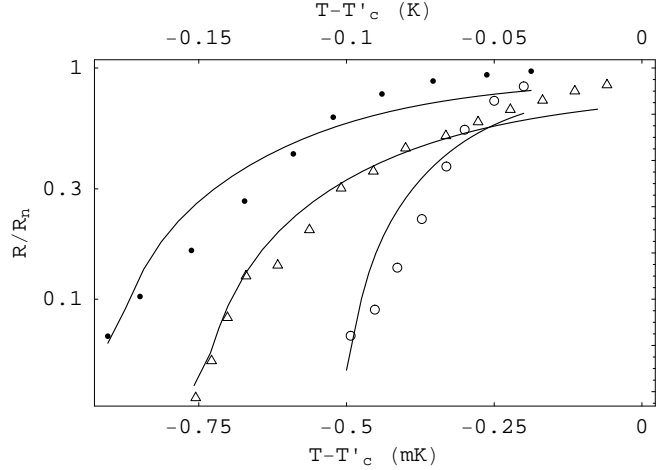}}%
\caption{\label{below} Resistances of indium microwires (upper scale) and a tin whisker (lower scale) as functions of temperature. $R$ is the resistance of the sample and $R_{\rm n}$ is its resistance in the normal state. $\triangle$ Sn, $w_0\sim 2\times 10^{-9}{\rm cm}^2$; $\bullet$ In, $w_0\sim 1.7\times 10^{-11}{\rm cm}^2$; $\circ$ In, $w_0\sim 5.2\times 10^{-11}{\rm cm}^2$.}
\end{figure}

We denote by $\Delta T=T'_c-T_c$ the difference between the the critical temperature estimated by the experimenters and that required by our model. We fitted the experimental data using our phenomenological Eqs. (\ref{empirpsi})--(\ref{modifiedAL}), with $\xi(0)$, $\xi_\beta$ and $\Delta T$ as free parameters. The results are shown in Figs.~\ref{leib}--\ref{below}; there is semiquantitative agreement, but the fitting functions are too concave. 

The values of the fitting parameters $\xi(0)$, $\xi_\beta$ and $\Delta T$ are displayed in Table~\ref{fits}. They cannot be taken too seriously, because it is possible to increase or decrease simultaneously the three parameters in a wide range without bringing about a clear difference in the goodness of the fit; moreover, the values of these parameters depend on the selected range. In all cases, $T_c$ is reasonably close to $T'_c$; for comparison, for the In microwires $T'_c$ differs by $\sim 800$mK from the critical temperature in bulk.

\begin{table*}
\caption{\label{fits}Parameters used in Eqs. (\ref{empirpsi}) and (\ref{sbal})} for the theoretical lines in Figs. \ref{leib}--\ref{below}. ``exp" stands for the values estimated in the experiments and ``fit" corresponds to the present fit. $\Delta T$ is the difference between the critical temperatures estimated in the experiments and those required here. The lengths $\xi(0)$ and $\xi_\beta$ are expressed in $\mu$m. $\xi(0)^{\rm exp}$ is not the same for the thin and the wide In wires, because they have different mean free paths.

\begin{ruledtabular}
\begin{tabular}{cccccc}
& $\Delta T$ (mK)& $\xi(0)^{\rm exp}$ & $\xi(0)^{\rm fit}$& $\xi_\beta^{\rm exp}$ & $\xi_\beta^{\rm fit}$\\
Al&22&0.15&1.1&3.1&5.8\\
In (thin)& 83&0.042&0.33&1.0&3\\In (wide)& 55&0.071&0.42&3.1&5.6\\
Sn&0.355&0.22&14&20&1870
\end{tabular}
\end{ruledtabular}
\end{table*}

$\xi(0)$ and $\xi_\beta$ enter our equations in two ways. One of them is through $\alpha '$, and in this case only their ratio $\xi_\beta/\xi(0)$ is of influence. In all cases, the experimental and the fitted ratio differ by less than half an order of magnitude, a factor that can be attributed to the coarse estimates, both in the experiment and in the fit. $\xi(0)$ or $\xi_\beta$ themselves enter through $K'$, which in turn depends on the assumption $\gamma=\pi\hbar^2/16k_BT'_cm\xi(0)^2$; in this case it seems that $\xi(0)^{\rm fit}$ and $\xi_\beta^{\rm fit}$ are too big. This might indicate that the theory for $\gamma$ is inappropriate and a more elaborate theory might be required \cite{Kramer}.

\subsection{A constriction}
We deal now with the case for which the present formalism has been designed, namely, a wire with nonuniform cross section. 
Denoting the arclength by $s$, we consider a wire such that its cross section is uniform in segments of length $N_{\rm cont}L/N$ at each extreme, assumes the value $w_{\rm ctr}\neq w_0$ at the position $s=s_{\rm ctr}$, $N_{\rm cont}L/N<s_{\rm ctr}<L-N_{\rm cont}L/N$, and is linear in $s$ for $N_{\rm cont}L/N\leq s\leq s_{\rm ctr}$ and $s_{\rm ctr}\leq s\leq L-N_{\rm cont}L/N$. We denote by $w_0$ the average cross section in the examined region $N_{\rm cont}L/N\leq s\leq L-N_{\rm cont}L/N$, i.e. the cross section in the contact regions has to be $2w_0-w_{\rm ctr}$. If $w_{\rm ctr}<w_0$ there is a constriction, and in the opposite case there is a widening. We still use the notations $\R=4e^2\gamma L/\hbar \sigma w_0$, $\xi_\beta=(\hbar^4w_0/4\beta m^2k_BT)^{1/3}$. In the discretized version we identify $w_i$ with the cross section at $s=(i-1/2)L/N$.

\subsubsection{Conductivity}
With these notations, Eq.~(\ref{oursig}) has to be replaced with
\be
\frac{\sigma '}{\sigma}=\frac{(N-2N_{\rm cont})\R \hbar Iw_0/e}{8N\gamma e\langle V\rangle (w_0-w_{\rm ctr})}
\log\frac{2w_0-w_{\rm ctr}}{w_{\rm ctr}}  -1 \;,
\label{ctrsig}
\ee
where $V$ is the potential drop excluding the initial and the final $N_{\rm cont}$ elements.

\begin{figure}
\scalebox{0.85}{\includegraphics{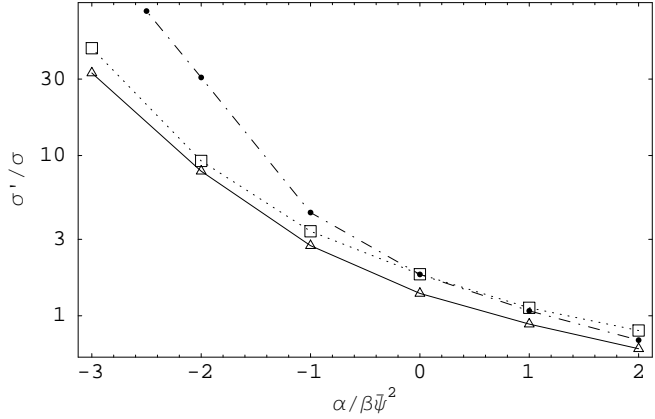}}%
\caption{\label{symmetric} Contribution of superconductivity to the conductivities of a uniform and two nonuniform wires, as functions of the temperature. 
$\L=10$, $\R=100$, $s_{\rm ctr}=0.5L$, $\xi_\beta =0.2L$, $I=0.1ek_BT/\hbar$, $N=21$, $N_{\rm cont}=3$, $\t_\beta =5\times 10^{-4}$, $N_{\rm av}=2\times 10^8$, $N_{\rm relax}=6\times 10^7$. For visibility, the symbols have been joined by straight segments. $\bullet$ uniform wire; $\triangle$ $w_{\rm ctr}=0.2w_0$; $\Box$ $w_{\rm ctr}=1.8w_0$.}
\end{figure}

Figure \ref{symmetric} compares the conductivities of a uniform and two nonuniform wires, as functions of the temperature. As the temperature decreases, the conductivities of the nonuniform wires increase at lower rates. An intuitive reason could be that in the thinner parts of the wire fluctuations are more effective in breaking superconductivity. Several evaluations of $\sigma '$ were repeated for different currents and we verified that our results correspond to the linear regime.

\subsubsection{Phase slips}
\begin{figure}
\scalebox{0.85}{\includegraphics{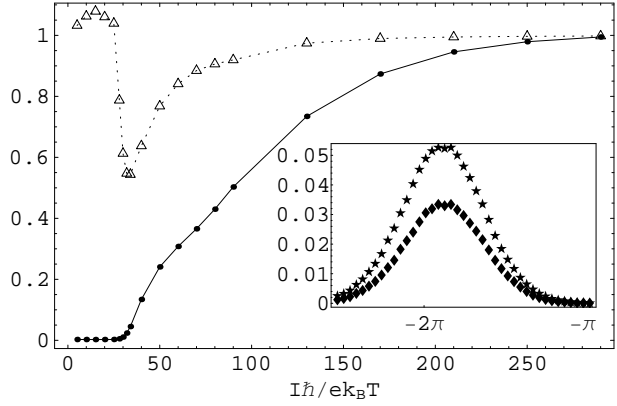} }%
\caption{\label{slipctr}Normalized voltage and voltage-order parameter correlation as functions of the current. $s_{\rm ctr}=0.5L$, $w_{\rm ctr}=0.2w_0$, $N=31$; the other parameters as in Fig.~\ref{vnormal} (for $I\le 25ek_BT/\hbar$ we took higher values of $N_{\rm av}$). $\bullet$ $V/V_{\rm normal}$; $\triangle$ $\langle V|\psi|_{\rm min}^2\rangle /\langle V\rangle \langle |\psi|_{\rm min}^2\rangle$. Insert: distribution of consecutive phase changes for $I\hbar /ek_BT=50$, as defined in Sec~\ref{VD}.  $\diamond$ Only includes phase changes restricted to the region above $-3\pi$; $\star$ phase changes below $-3\pi$ were added after being brought into the region between $-3\pi$ and $-\pi$, by addition of the appropriate integer multiple of $2\pi$.  }
\end{figure}

We repeat our study of Sec.~\ref{High}, but this time the wire has a constriction in the middle. The results are shown in Fig.~\ref{slipctr}. Our results for $\langle V|\psi|_{\rm min}^2\rangle /\langle V\rangle \langle |\psi|_{\rm min}^2\rangle$ indicate that the constriction enhances the influence of fluctuations and the phase-slip region is both enlarged and shifted to lower currents. A shift to lower currents was also found in \cite{MichotteB69}. A curious effect is that, for currents slightly below this region, the voltage and the minimum size of the order parameter are positively correlated.

The insert shows the probability density for consecutive phase change, for a current in the phase-slip region. As expected, the constriction promotes the occurrence of phase slips: the probability density near $-2\pi$ is much larger than in the case of a uniform wire, and also drops faster away from $-2\pi$. Also, when the order parameter vanishes during a phase slip, it now happens almost exclusively at the constriction.

\subsubsection{Charge accumulation}
As a last test for statistical mechanics, we disconnect the current source and connect a capacitor $C$ between the points $s=0$ and $s=s_C$. This capacitor will be charged with a charge $Q$ that fluctuates in time. The region $0\le s\le s_C$ will be forced to sustain a potential drop $Q/C$ and the current $I_1$ in this this region will be obtained as a Lagrange multiplier, as in Sec.~\ref{non}. The current in the region $s_C\le s\le L$ has to vanish. During each step, the charge in the capacitor increases by the amount $I_1\tau$.

From the point of view of the capacitor, the wire is just a heat reservoir. Therefore, writing $\tilde Q=Q/\sqrt{k_BTC}$, the probability density for the charge will be proportional to $\exp (-\tilde Q^2/2)$. From here, we expect $\langle |\tilde Q|^n\rangle =2^{n/2}\Gamma ((n+1)/2)/\sqrt{\pi}$. Table~\ref{capacitor} compares this prediction with the results obtained with our method.

\begin{table*}
\caption{\label{capacitor} Average value of $|\tilde Q|^n$ according to statistical mechanics and to the method developed here. $s_{\rm ctr}=0.36L$, $s_C=0.48L$, $C=0.01\gamma^2 e^2/k_BT$; the other relevant parameters are as in Fig.~\ref{symmetric}.}

\begin{ruledtabular}
\begin{tabular}{ccccc}
 $n$ & 1 & 2& 3 & 4\\
$2^{n/2}\Gamma ((n+1)/2)/\sqrt{\pi}$&0.80&1&1.60&3\\
our simulation& 0.81&1.02&1.65&3.14
\end{tabular}
\end{ruledtabular}
\end{table*}

\section{Summary}
We have developed a formalism that takes into account the influence of thermal fluctuations in a one-dimensional superconductor described by the TDGL model. Our method incorporates the fluctuation-dissipation theorem in the form required for integration of a discretized stochastic partial differential equation by means of Euler iterations.
 We have pointed out the relationship between the size of the fluctuations and the volume of the elements into which the sample is divided; to my knowledge, this relationship has not been previously addressed in the literature. We have verified that our method complies with statistical mechanics.

Using a gauged order parameter ($\ps$) rather than the raw order parameter ($\psi $), we obtain a formalism in which the electromagnetic potential is not needed in the macroscopic evolution. Special care is required in the use of this transformation; although the thermodynamic potential is more naturally expressed in terms of $\ps$, the variable that permits direct application of the fluctuation--dissipation theorem is $\psi $.

As an application, we have evaluated the contribution of incipient superconductivity to the electrical conductivity. Significantly above the critical temperature, we recover the Aslamazov--Larkin result. We have focused on the regime immediately below the critical temperature. We compared our results with those of old experiments in this regime, and the agreement is poor. The lack of agreement was expected and does not arise from the way we have treated fluctuations, but rather from TDGL itself; no available theory reproduces the experimental data in this regime (although the Hartree approximation has been successful in situations in which fluctuations may be regarded as effectively one-dimensional \cite{Nikulov}).

Our method enables us to deal with wires with nonuniform cross section. In particular, we studied the influence of fluctuations and constrictions on the presence and behavior of phase slips. When a constriction is present, phase slips appear at lower currents.

Our formalism can be easily generalized to more complex situations, such as multiply connected topologies where non-integer fluxes are enclosed, time-dependent applied voltages or currents, moderate variations in temperature, or different evolution equations that can be cast in the form of dissipative equations. Generalization to models that are not purely dissipative (e.g. \cite{don}) is not immediate.

\begin{acknowledgments}
This work has been supported by the Israel Science Foundation under grant 4/03-11.7. I am grateful to Jens-Peer Kuska and Aaron Suggs for teaching me how to include fonts in the figures and to Alexey Nikulov, Jacob Rubinstein and Boris Shapiro for useful comments.
\end{acknowledgments}

\appendix
\section{Numeric Considerations\label{AA}}
A well-known difficulty in the solution of partial differential equations by means of Euler iterations is that convergence requires quadratically small time steps as the computational grid becomes dense. The standard way for overcoming this difficulty is the Crank-Nicolson implicit scheme. However, we did not find that this scheme is useful when stochastic noise is present. Our general strategy was to perform analytically the most important parts of the iteration and then add a simple Euler iteration for the remaining part. Our final algorithm was remarkably stable and enabled us to vary practically all the parameters in our model over several orders of magnitude.

\subsection{Treatment of the inter-cell interaction\label{external}}
For a dense computational grid $\tilde\xi\gg 1$ and the inter-cell interaction becomes the leading term in Eq.~(\ref{GGLtil}). We treat this term separately. Regarding $\bm{\psi}=(\ps_1,\dots ,\ps_N)$ as a vector, we have to leading term
\be
\frac{d\bm{\psi}}{dt}={\bf M}\bm{\psi} \;,
\label{veceq}
\ee
where the matrix ${\bf M}$ has the elements $-2\alpha\xiq/\gamma \hbar$ along the diagonal, $\alpha\xiq/\gamma \hbar$ when the column and the row number differ by 1, and zeros everywhere else. The first and the last rows of ${\bf M}$ depend on the boundary conditions; for instance, for periodic boundary conditions $M_{1N}=M_{N1}=\alpha\xiq/\gamma \hbar$, whereas for Neumann conditions $M_{1N}=M_{N1}=0$ and $M_{11}=M_{NN}=-\alpha\xiq/\gamma \hbar$.
Ignoring the other contributions, Eq.~(\ref{veceq}) can be integrated to give
\be
\bm{\psi}(t+\tau )=\exp({\bf M}\tau )\bm{\psi}(t) \;.
\label{matrixev}
\ee
${\bf M}$ has eigenvalues that are either negative or zero, so that the largest eigenvalue of $\exp({\bf M}\tau )$ is 1.

We can therefore follow the evolution of $\ps_k$ by means of steps that consist of three stages. The first stage is described by Eq.~(\ref{matrixev}) and the other two are as described at the end of Sec. \ref{GI}. When considering the contribution of the first term in Eq.~(\ref{29}), the leading term is dropped, since it has already been taken into account. The matrix $\exp({\bf M}\tau )$ is the same for every step, so that it has to be calculated just once.

\subsection{Treatment of the condensation-energy term}
The evolution equation (\ref{Dpsibeta}) contains two contributions to $\Delta_{\rm mac}\ps_{\beta k}$: an ``internal" contribution $-\t_\beta (\alpha '+|\ps_{\beta k}|^2)\ps_{\beta k}$, that will be considered here, and an ``external" contribution due to the interaction with the neighboring elements $k\pm 1$, that was considered above. 

If only the internal contribution is considered, the evolution equation can be integrated and gives $\ps_{\beta k}(t+\tau )=(\alpha '/[(\alpha '+|\ps_{\beta k}(t)|^2)\exp(2\alpha '\t_\beta )-|\ps_{\beta k}(t)|^2] )^{1/2}\ps_{\beta k}(t)$. Substituting this expression by a computationally less expensive Pad\'e approximant we obtain the algorithm
\be
\ps_{\beta k}\leftarrow\frac{2+(|\ps_{\beta k}|^2-\alpha ')\t_\beta}{2+(3|\ps_{\beta k}|^2+\alpha ')\t_\beta}\ps_{\beta k}
\label{algointern}
\ee
that takes into account the internal contribution.

\subsection{Step size}
Let us first consider the case $\beta =0$. For the validity of the algorithm (\ref{algo}) we require $\tau \ll k_BT/\Gamma_\psi (\partial G/\partial u_k)^2$. We make the estimates $\partial G/\partial u_k\sim \alpha w_0L\xiq |\psi_k|/N$ and $|\psi_k|\sim\sqrt{N/\AL w_0L}$. On the other hand, the fluctuations of $I_{\rm SC}$, that were estimated in Sec.~\ref{vol-cur1}, have to be small compared to $\langle I_{\rm SC}\rangle$. Therefore, $\t$ has to be in the range
\be
400\xi/N_{\rm av}L\ll\t\ll\tilde\xi^{-4} \;.
\label{rangepsi}
\ee
Similarly, we require $\tau \ll k_BT/\Gamma_A (\partial G/\partial \A_k)^2$ and $\langle I_{SC}\rangle$ has to be large compared to the Johnson noise. From here we obtain
\be
800\AL/N_{\rm av}\R\ll\t\ll N\AL/\R\tilde\xi^4 \;.
\label{rangeA}
\ee
In addition, we require $N_{\rm relax} \t\gg 1$.

Our analysis for $\beta\neq 0$ is similar to that leading to Eqs. (\ref{rangepsi})-(\ref{rangeA}), but now we use the estimate $|\psi_k|\sim\bar\psi$.
Eq.~(\ref{44}) is still qualitatively true, as shown in Fig.~\ref{psisq}. With these estimates we obtain
\begin{eqnarray}
100 \xi_\beta /N_{\rm av}L & \ll & \t_\beta \ll  (L/N\xi_\beta)^3 \;,\label{rangepsiB}\\
800L^2/N_{\rm av}\R\xi_\beta ^2\L^2 & \ll & \t_\beta  \ll  L^4/N\R\xi_\beta^4\L^2 \;. \label{rangeAB}
\end{eqnarray}

\section{Average supercurrent}
In this appendix we estimate the characteristic size of supercurrents in the absence of voltage. Let us consider a wire with uniform cross section with periodic boundary conditions. The average of the supercurrent along the wire is
\be
I_{\rm SC}=\frac{4e\alpha\xiq Lw_0}{\hbar N^2}\sum_{j=1}^N {\rm Im}\left[\ps_{j}^*\ps_{j-1}\right] \;,
\label{sc1}
\ee
which can also be written as
\be
I_{\rm SC}=-\frac{4e\alpha\xiq Lw_0}{\hbar N}\sum_{n=0}^{N-1} |\tilde\varphi_n|^2\sin\frac{2\pi n}{N} \;.
\label{sc2}
\ee

The ensemble average of $I_{\rm SC}$ is zero. Its variance is obtained from the square of Eq.~(\ref{sc2}),
\be
I_{\rm SC}^2=\frac{(4e\alpha\xiq Lw_0)^2}{(\hbar N)^2}\sum_{n,n'}|\tilde\varphi_n|^2|\tilde\varphi_{n'}|^2\sin\frac{2\pi n}{N}\sin\frac{2\pi n'}{N} \;.
\label{scsq}
\ee
The expression for the variance is simplified by substraction of the null quantity $\langle I_{\rm SC}\rangle^2$ and noting that for $n\neq n'$ $\langle|\tilde\varphi_n|^2|\tilde\varphi_{n'}|^2\rangle=\langle|\tilde\varphi_n|^2\rangle\langle|\tilde\varphi_{n'}|^2\rangle$, whereas $\langle|\tilde\varphi_n|^4\rangle=2\langle|\tilde\varphi_n|^2\rangle^2$. We obtain
\be
\langle I_{\rm SC}^2\rangle=\frac{(4e\xiq k_BT)^2}{(\hbar N)^2}\sum_{n=0}^{N-1}\frac{\sin^2(2\pi n/N)}{(1+4\xiq \sin^2(\pi n/N))^2} \;.
\label{varsc}
\ee

For $N\gg 1$, the sum can be replaced with an integral; if in addition we assume $\tilde\xi\gg 1$ and keep only the leading term, Eq.~(\ref{varsc}) becomes
\be
\langle I_{\rm SC}^2\rangle=\frac{4e^2 k_B^2T^2\xi}{\hbar^2 L} \;.
\label{44}
\ee

\end{document}